%% file: acl_latex.tex
\documentclass[11pt]{article}

\usepackage{times}
\usepackage{latexsym}

\usepackage[T1]{fontenc}
\usepackage[final]{acl}
\usepackage{times}
\usepackage{latexsym}

\usepackage[T1]{fontenc}
\usepackage[utf8]{inputenc}

\usepackage{microtype}

\usepackage{inconsolata}

\input{common.tex}

\title{\sysname: Toward Functionally Correct and Secure Code Generation\\via Multi-Agent Collaboration}
\author{
  Miseon Yu \\
  Seoul National University \\
  Seoul, Republic of Korea \\
  \texttt{altjs543@snu.ac.kr}
  \And
  Jaehoon Choi \\
  Seoul National University \\
  Seoul, Republic of Korea \\
  \texttt{qkzkf011113@gmail.com}
  \AND
  Younghan Lee\thanks{Corresponding authors: Younghan Lee (\texttt{yhlee@sungshin.ac.kr}) and Yunheung Paek (\texttt{ypaek@snu.ac.kr}).} \\
  Sungshin Women's University \\
  Seoul, Republic of Korea \\
  \texttt{yhlee@sungshin.ac.kr}
  \And
  Yunheung Paek\footnotemark[1] \\
  Seoul National University \\
  Seoul, Republic of Korea \\
  \texttt{ypaek@snu.ac.kr}
}

\begin{document}
\maketitle
\begin{abstract}
% Abstract: ≤200 words
Despite their strong ability to generate code, large language models often fail to produce secure code, as their outputs frequently contain security vulnerabilities. Secure code generation is inherently challenging because it requires solving a multi-objective problem: functional correctness and security. \final{Existing approaches address this challenge by injecting external security knowledge or by using agentic feedback and iterative refinement. However, guideline retrieval often leaves the generator to translate generic advice into task-specific secure implementations, while shared-dialogue multi-agent feedback can blur role boundaries and suffer from context bloat.}
We present \sysname, a multi-agent framework that integrates \textit{planning, security analysis, code synthesis} and \textit{refinement} to jointly optimize security and functionality. A planner constructs a step-by-step plan to satisfy functional requirements. A security advisor identifies likely CWEs and synthesizes task-specific guidelines, \final{a coder then generates code grounded in these artifacts, and a reviewer issues perspective-separated feedback. Rather than sharing full dialogue histories, each agent receives only structured artifacts from upstream stages, enforcing role specialization and reducing uncontrolled context growth.} 
\yhlee{On CWEval and BaxBench, \sysname\ improves F\&S@1 over direct prompting by 19.61 and 10.57 percentage points (pp) on average, respectively.}
%(19.33+20.17+19.33+22.69+17.65+17.65)/6
%(20.17+20.17+19.33+22.69+17.65+17.65)/6
%((31.12-21.17)+(21.94-10.76))/2 
%iEvaluated on six LLMs and eight programming languages using CWEval and BaxBench, \sysname\ achieves an average improvement of \finalsnf\ in the combined functionality-and-security score.
\end{abstract}

\input{sections/1_Introduction}

\input{sections/2_Background}
\input{sections/3_Method}
\input{sections/4_Experiment}

\input{sections/5_Results}

\input{sections/6_Ablation}
\input{sections/7_Conclusion}
\clearpage
\input{sections/Limitations}
\input{sections/Ethics}
% ~\ref{sec:appendix}
%\section*{Acknowledgments}


% \bibliography{custom}
\bibliography{custom}
\clearpage
\appendix

\input{sections/Appendix}

\end{document}

%% file: common.tex
\usepackage{amsmath, amsopn}
\usepackage{booktabs, array, makecell, hhline, multirow}
\usepackage{graphicx, subcaption}
\usepackage{xfrac, nicefrac}
\usepackage{adjustbox}
\usepackage{microtype, xspace, url}
\usepackage{enumitem}
\usepackage{siunitx, fp}
\usepackage{balance, multicol}
\usepackage{siunitx}
\definecolor{cadmiumgreen}{rgb}{0.0, 0.42, 0.24}
\definecolor{forestgreen}{HTML}{009B55}
\usepackage[table]{xcolor}
\usepackage{pifont}
\usepackage{listings}   
\usepackage{amsmath}
\usepackage{amssymb}
\newcommand{\incmark}{\textcolor{green!50!black}{\ding{51}}} 
\newcommand{\inxmark}{\textcolor{red!70!black}{\ding{55}}}  
\newif\ifhighlight
\highlightfalse

\newif\ifrevision
\revisionfalse     %  red 

\newif\iffinal
\finalfalse         % teal 
\newif\ifreplace
\replacefalse      % blue 

\newcommand{\revision}[1]{%
  \ifrevision
    \textcolor{magenta}{#1}%
  \else
    #1%
  \fi
}

\newcommand{\final}[1]{%
  \iffinal
    \textcolor{red}{#1}%
  \else
    #1%
  \fi
}

\newcommand{\replace}[1]{%
  \ifreplace
    \textcolor{blue}{#1}%
  \else
    #1%
  \fi
}

\newcommand{\yhlee}[1]{\textcolor{black}{#1}}

\newcommand{\sysname}{\textsc{MACGen}}

%% file: sections/1_Introduction.tex
\begin{figure}[!ht]
\centering   
\includegraphics[width=0.98\linewidth]{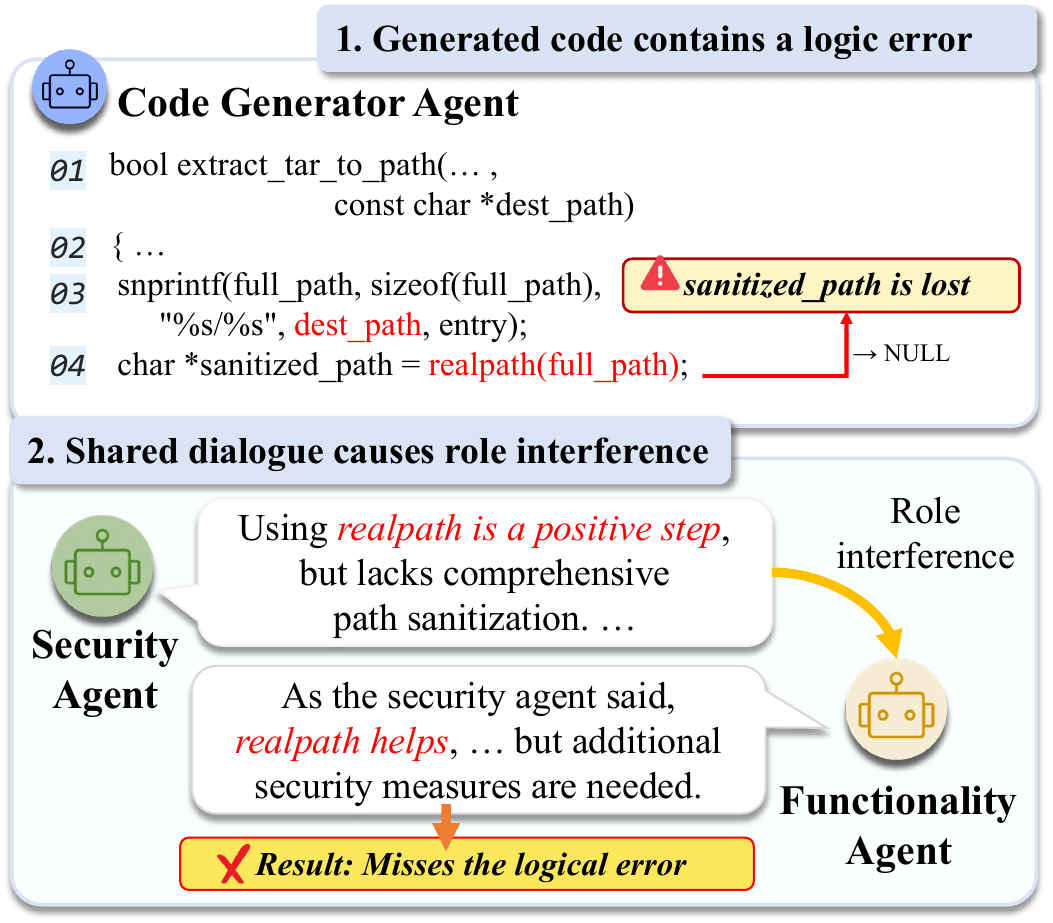}
    \caption{
\final{Example of role interference. The generated code calls \texttt{realpath} on a non-existent destination path, causing \texttt{sanitized\_path} to become \texttt{NULL} rather than preserving the intended extraction path. After observing the security critic's feedback in the shared-dialogue setting, the functionality critic shifts toward security-oriented feedback rather than independently identifying this logic error.}
}
\label{fig:motivating}
\end{figure}

\section{Introduction}
Large language models (LLMs) demonstrate strong code-generation capabilities and are rapidly being adopted in practice~\citep{githubcopilot,github2024aiwave}. Empirical studies, however, reveal that LLM-generated code often contains a significant number of vulnerabilities~\citep{asleep, khoury2023secure, cyberseceval, mou-etal-2025-really}, including many listed in the MITRE CWE\footnote{The Common Weakness Enumeration (CWE) is a public catalog of software weaknesses comprising over 900 types.~\url{https://cwe.mitre.org/news/archives/news2025.html}}
 Top-25~\citep{mitre-cwe-top25-2024}.
 
\yhlee{To mitigate these vulnerabilities, recent work has framed secure code generation as the task of producing code that satisfies the requested functionality while avoiding implementation choices that introduce security weaknesses. Unlike a post-generation filtering problem, secure code generation requires security decisions to be made within the implementation itself, where they can affect whether the program still satisfies the intended behavior.}
\final{Such dual-objective setting increases task complexity, as secure implementations require additional decisions such as input validation, safer API selection, or permission checks. Moreover, the security risks relevant to a task are often underspecified in the prompt, requiring the model to infer applicable threats and translate them into concrete, functionally consistent code choices. Empirically, security-oriented prompting, which directly instructs the model to produce secure code, can reduce vulnerabilities but often compromises functional correctness~\citep{prompt_tech_scg,balancing,rethinkscg}. The expanded reasoning burden imposed by dual objectives is difficult to resolve reliably through prompting alone.}

\final{One line of work mitigates this challenge by injecting external security knowledge into the generation context, such as secure code examples or security guidelines~\citep{zhang-etal-2024-seccoder,codeguader,rescue}. More recently, these approaches have been refined by selecting concise and task-relevant guidelines to better preserve functionality~\citep{rescue}. However, such guidelines are inherently generic and lack awareness of task-specific context such as variable names, function signatures, or control flow. Similarly, secure code examples are language-specific and costly to collect, limiting their practical coverage. As a result, even with retrieved knowledge in context, the generator must still determine how generic guidance applies to the specific task and translate it into secure, functional code. Thus, retrieval narrows the security-knowledge gap but leaves task-specific adaptation unresolved at generation time.}

\final{Another line of work addresses this reasoning burden through agentic critique. \citet{le2024indict} propose an internal-dialogue framework in which safety and helpfulness critic agents analyze generated code from their respective perspectives. By assigning each concern to a distinct critic, this approach makes functional and security-related evaluation criteria explicit. However, shared-dialogue designs may blur these role boundaries when critics observe each other’s intermediate reasoning, potentially leading to role interference in which critics produce overlapping feedback or gradually drift from their assigned perspective (Figure~\ref{fig:motivating}). Because role separation is enforced only through prompt instructions rather than by the interface itself, such designs rely on accumulated dialogue to coordinate critiques. This increases token cost and latency while exposing agents to long-context degradation as the dialogue history grows~\citep{liu-etal-2024-lost,liu2024agentbench}.}

\final{These considerations suggest that effective secure code generation requires not just more knowledge or more feedback, but a clearer separation of what each stage needs to reason about. Each stage should receive only the information it needs and pass a compact artifact to the next, so that role boundaries are enforced by the interface rather than by prompts alone. We instantiate this principle as \sysname, a multi-agent collaboration system for secure code generation. Inspired by the Secure Software Development Lifecycle (SSDLC)~\citep{howard2006sdl}, which emphasizes incorporating security throughout development, \sysname\ decomposes secure code generation into planning, security analysis, synthesis, and refinement, each handled by a dedicated agent.}

\yhlee{Specifically, a planner first establishes a functional plan, analogous to clarifying requirements. A code generator then produces draft code from this plan and a security advisor then analyzes the task, plan, and draft code to identify threats and generate task-specific guidelines that cover relevant CWE risks, serving a role similar to incorporating security during design and bridging the gap between generic retrieved advice and the concrete coding task. Subsequently, a code generator implements secure code guided by both the plan and security constraints. Finally, a reviewer supports the verification phase by issuing actionable, perspective-separated feedback routed back to the coder for targeted refinement. By resolving what to implement and how to secure it before code synthesis, \sysname\ helps the coder produce code that satisfies both functional and security requirements from the outset. Unlike prior multi‑agent approaches that rely on shared dialogue, our framework enforces role specialization efficiently through artifact‑only interfaces.}

We conduct a comprehensive evaluation of \sysname\ using two complementary benchmarks, CWEval~\citep{cwevalbench} and BaxBench~\citep{baxbench}. \yhlee{On CWEval, across six LLMs and five programming languages, \sysname\ improves F\&S@1 over direct prompting by 19.61 percentage points (pp) on average. On BaxBench, evaluated with GPT-4o and GPT-4o-mini across six backend languages, \sysname\ improves F\&S@1 by 9.95pp and 11.18pp, respectively.}

\begin{itemize}
    \item \yhlee{We introduce \sysname, a multi-agent framework for functionally correct and secure code generation that decomposes generation into planning, security analysis, code synthesis, and refinement.}
    
    \item \yhlee{We show that artifact-only interfaces provide an effective coordination structure for specialized agents, by reducing unnecessary context sharing between roles.}
    
    \item \yhlee{We evaluate \sysname\ ~on CWEval and BaxBench across multiple LLMs and languages, reporting improved joint functionality-security performance.}~\footnote{All source code, evaluation logs, and prompts are available at \url{https://github.com/Mishuni/macgen-secure-code.git}}.
    %We conduct extensive experiments exhibiting significant performance gains across multiple LLMs and programming languages
\end{itemize}

%% file: sections/2_Background.tex
\begin{figure*}[!ht]
\centering   
\includegraphics[width=0.99\linewidth]{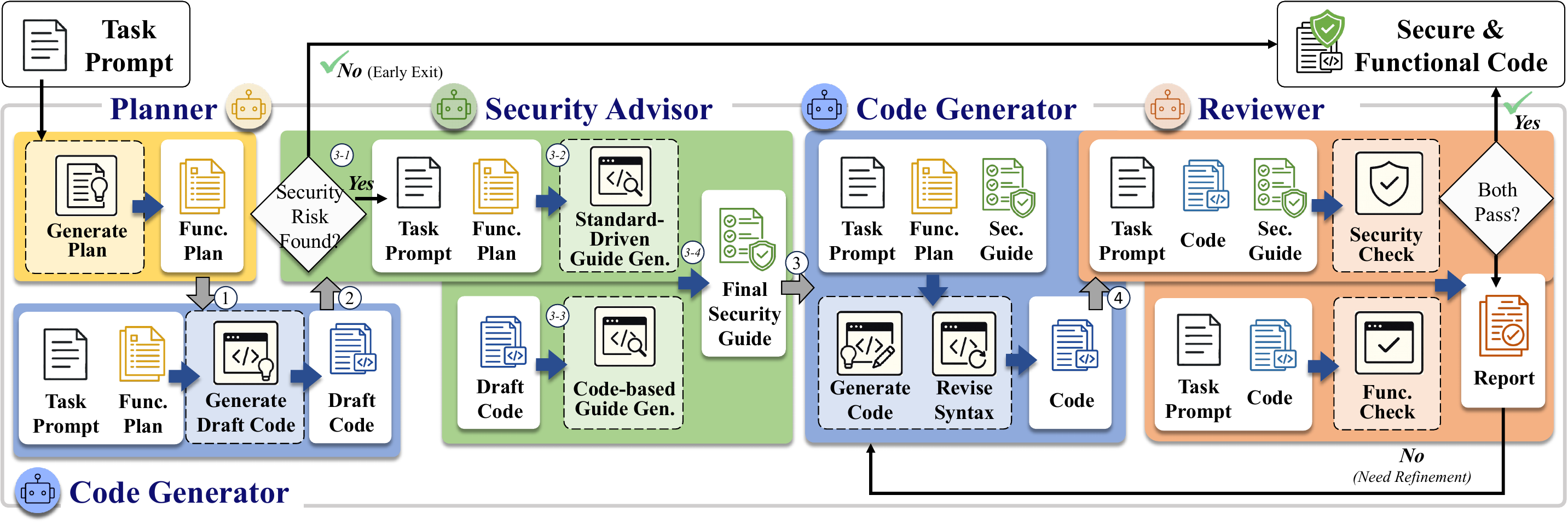}
    \caption{Overview of \sysname: The framework consists of four specialized agents—Planner, Security Advisor, Code Generator, and Reviewer.}
    \label{fig:main}
\end{figure*}
\section{Related Works}

% For introducing Planning 
% Planning based Code generation ??
\subsection{Multi-Agents for Code Generation}
 \label{rel:code_gen}

Multi-agent strategies have shown potential for improving code quality through structured task decomposition~\citep{li2023camel, hong2024metagpt, qian2023chatdev, islam-etal-2024-mapcoder, islam-etal-2025-codesim}.
Especially, \citet{islam-etal-2025-codesim} introduces \textsc{CodeSIM}, \revision{which} significantly enhances code quality by leveraging planning techniques inspired by human problem-solving processes. It generates a step-by-step implementation plan, which is then simulated using synthesized input/output samples to verify correctness. This plan is iteratively refined \revision{until the simulation succeeds}. Despite these advances, such planning-based and human problem-solving approaches have yet to be fully utilized in the domain of secure code generation. \revision{\sysname\ extends this line of work by applying planning techniques to secure code generation.}

\subsection{Secure Code Generation}
\final{\textit{Model adaptation} improves security via security-centric prefix-/instruction-tuning, localized optimization, or internal feature control~\citep{sven,safecoder,CoSec,hasan-etal-2025-teaching,deepguard,elhusseini2026repairing}. However, such approaches require access to model parameters or internals,
limiting applicability when only inference-time interaction is available.} \textit{Prompting/Reflection} can steer LLMs toward safer code at inference time~\citep{nazzal2024promsec,prompt_tech_scg,bruni2025benchmarking}; for example, \citet{prompt_tech_scg} show that multi-turn Recursive Criticism and Improvement (RCI) significantly reduces security weaknesses. \final{\textit{Knowledge Injection} augments the generation context with external security knowledge, such as secure code examples or guidelines~\citep{reflexgen,zhang-etal-2024-seccoder,secguide,codeguader,rescue}.
\textit{Agentic Critique} uses specialized agents to review generated code for security~\citep{autosafecoder} or for both functional and security perspectives~\citep{le2024indict}. Collectively, these inference-time approaches treat security as added knowledge or post-generation feedback, rather than integrating it throughout the development process as SSDLC principles prescribe.}

%% file: sections/3_Method.tex
\section{\sysname\ Design}

\subsection{Problem Definition}
\label{overall:pd}
Our goal is to generate source code that is syntactically valid, functionally correct, and secure. We approach this as a conditional generation task where a multi-agent system produces a code sequence~$Y$ from a prompt~$X$. This input~$X$ can consist of a natural language instruction, a code prefix for completion, or a combination of both.

\subsection{Overall Workflow}
\replace{\sysname\ is a role-specialized multi-agent framework composed of four agents: a Planner, a Security Advisor, a Code Generator, and a Reviewer. An overview of the framework is displayed in Fig~\ref{fig:main}.} \revision{The process begins with the Planner establishing a functional plan, which the Code Generator uses to produce the draft code (\textcircled{1}–\textcircled{2} in Fig~\ref{fig:main}). The Security Advisor then analyzes the draft, either triggering an early exit or performing standards-driven reasoning via a security Knowledge Base (KB) to synthesize guidelines (\textcircled{3}). Finally, the Code Generator implements the final code, followed by iterative verification from the Reviewer (\textcircled{4}).}

\subsection{Building a Standards-driven Security Knowledge Base}
\label{sec:kb}
To ensure the normative correctness of security reasoning, we construct a standards-grounded KB from official code security standards such as CERT and OWASP~\citep{cert,owasp_asvs}. This pre-constructed KB serves as the foundation for the retrieval mechanism, enabling the Security Advisor to perform standards-grounded reasoning.

Raw text parsed from heterogeneous sources (e.g., PDF, Markdown) often contains structural noise and formatting inconsistencies that hinder effective semantic retrieval. To address this, we leverage an LLM-based refinement process that normalizes each guideline into a compact, high-density schema consisting of four fields (WHAT, WHY, HOW, and EXAMPLE). The prompt used for this refinement is provided in Figure~\ref{fig:prompt_secadv1}.

To convert these structured guidelines into a searchable semantic space, each guide $g_i$
is encoded into a vector $\mathbf{v}_i \in \mathbb{R}^d$ using an embedding function $E(\cdot)$.
During inference, the Security Advisor formulates a set of task-specific queries
$\mathcal{Q} = \{q_1, \dots, q_m\}$ derived from the identified CWE groups. For each query $q_j \in \mathcal{Q}$, the retrieval process selects the top-$K$ most relevant guidelines according to cosine similarity:
\begin{equation}
G^*(q_j) = \mathop{\mathrm{arg\,max}^{(K)}}_{g_i \in \mathrm{KB}} \; \mathrm{sim}\big(E(q_j), E(g_i)\big)
\end{equation}
\revision{where $\operatorname{arg\,max}^{(K)}$ denotes an operator that returns the set of
$K$ elements with the highest similarity scores, and
$\mathrm{sim}(\cdot,\cdot)$ represents cosine similarity.
The final candidate set $G$ is obtained by the union of retrieved guidelines across all queries in $\mathcal{Q}$.}
% G^*(q_j) = \arg\max_{g_i \in \mathrm{KB}} \; \mathrm{sim}\big(E(q_j), E(g_i)\big)

\subsection{Role-specific Agents}
\revision{This subsection details the roles and internal logic of each agent. The complete prompts for all agents are provided in Appendix~\ref{apx:prompts}.}
% Note that the complete prompts for agents discussed in this subsection are provided in Appendix~\ref{apx:prompts}.

\subsubsection{Planner}
\final{Inspired by prior work~\citep{islam-etal-2025-codesim}, the Planner agent generates a concise, high-level functional plan for each task. This plan serves dual purposes. It guides the Code Generator toward a functionally correct implementation, and provides task-specific context that enables the Security Advisor to identify relevant attack surfaces more precisely.}
\begin{figure}[!ht]
\centering   
\includegraphics[width=0.90\linewidth]{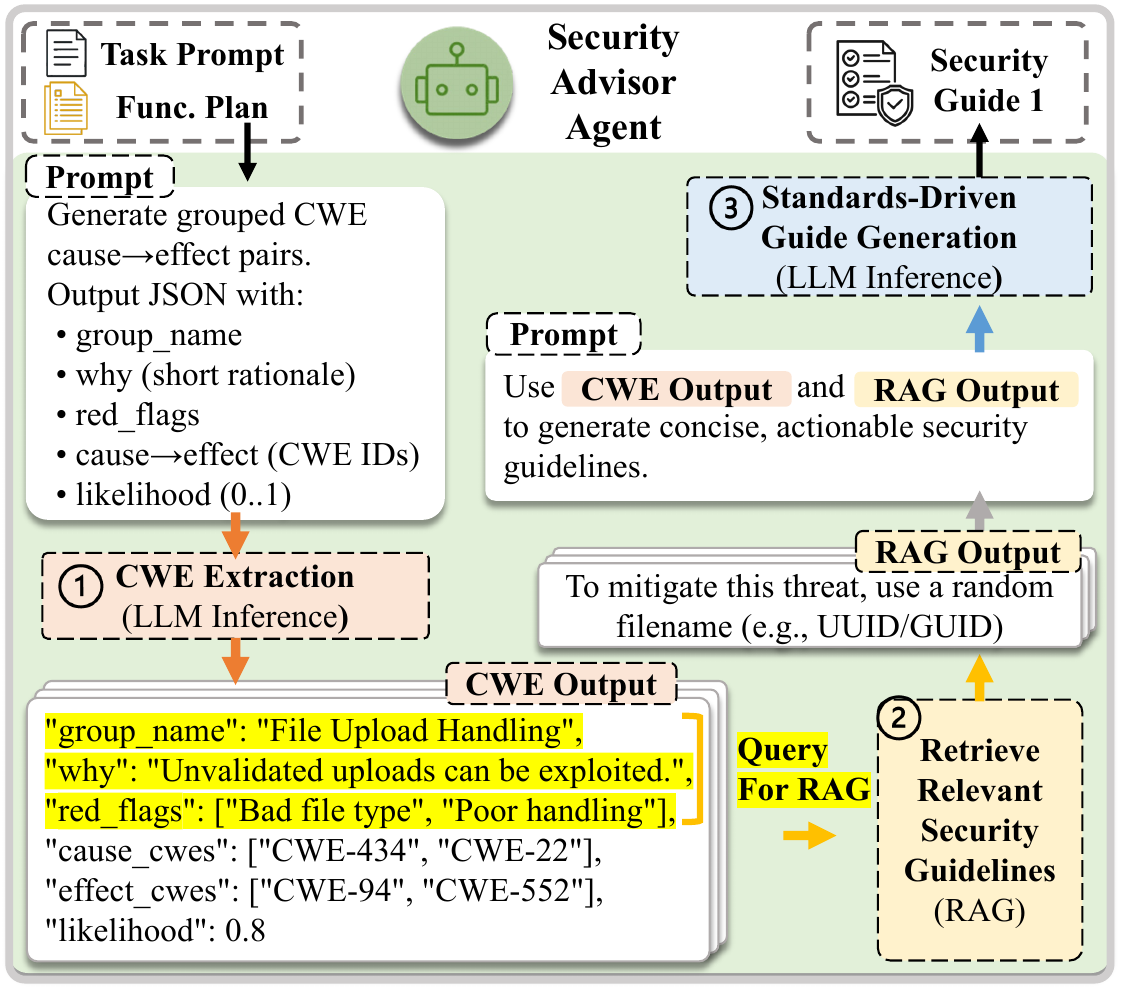}
    \caption{Detailed visualization of the standards-driven guideline generation process (Step 3-2 in Fig~\ref{fig:main}).}
    \label{fig:secadv}
\end{figure}
\subsubsection{Security Advisor}
The Security Advisor follows a multi-stage pipeline designed for both inference efficiency and analytical depth. The process begins with an early-exit triage, where the advisor inspects the task prompt and the initial draft code to identify potential attack surfaces under pre-defined categories. If no risks are detected (e.g., for simple algorithmic or purely functional tasks), the system executes an \textit{early exit} to bypass unnecessary retrieval and minimize computational overhead.

For tasks requiring deeper inspection, the advisor synthesizes task-specific guidelines through a three-step process. First, following the workflow in Figure~\ref{fig:secadv}, it examines the task prompt and the functional plan to identify potential security risks, which are represented as corresponding CWE groups (e.g., insecure file path handling). These groups are then used to formulate targeted search queries for retrieving relevant security guidelines from the pre-constructed KB. Based on the retrieved context, the advisor generates the first set of standards-driven security guidelines.

Subsequently, the advisor performs a code-based guide generation by inspecting the draft code. This step is crucial for detecting insecure coding patterns typical of LLM outputs that may not be apparent from the high-level plan alone. Based on this code-level analysis, the second set of guidelines is generated. Finally, the advisor concatenates both sets and validates them against the functional requirements to ensure no conflicts exist. This consolidation guarantees that the security requirements are both comprehensive and compatible with the task’s functionality.

\subsubsection{Code Generator}
The Code Generator agent serves two distinct roles within the pipeline.
First, it produces an initial draft code based on the task prompt and the functional plan, which is subsequently analyzed by the Security Advisor. Second, guided by the functional plan and the security guidelines, the agent generates the executable code, with both components acting as dual constraints on the output. After generation, the agent iteratively checks the generated code using a syntax checker and performs self-refinement until the code is syntactically valid.

\subsubsection{Reviewer}
The Reviewer agent conducts a structured code review \final{through two perspective-separated checks. It verifies that the implementation accomplishes the intended functionality and that all security guidelines have been correctly met. If it uncovers any functional errors (e.g., incorrect logic) or residual security issues (e.g., missing input sanitization, use of risky functions), it provides targeted feedback.} This feedback is passed to the Code Generator, which revises the implementation as needed. The iterative refinement cycle continues until the Reviewer approves the code as both functionally correct and secure, or the maximum number of iterations is reached.

%% file: sections/4_Experiment.tex
\section{Experimental Setup}\label{sec:experiments}
In this section, we describe the experimental setup, including benchmarks, models, baselines, implementation details, and evaluation metrics. More details are provided in Appendix~\ref{apx:experiments_setup}.
\subsection{Benchmarks}
We evaluate our approach using two security-oriented code generation benchmarks.
\final{First, we employ CWEval~\citep{cwevalbench}, an outcome-driven evaluation benchmark comprising 119 code completion tasks across five programming languages (C, C++, Python, JavaScript, and Go).
Second, we use BaxBench~\citep{baxbench}, a benchmark for generating backend applications in realistic and diverse environments. It contains 392 tasks across 28 backend scenarios and 14 frameworks in six programming languages (Go, JavaScript, PHP, Python, Ruby, and Rust), spans both single- and multi-file application settings.}

\input{tables/main/exp1_cweval_revision}

\subsection{Baselines \& LLM Models}
To evaluate the effectiveness of our framework, we compare it against five baselines: Direct Prompting where LLMs generate code $Y$ directly from the task prompt $X$, \textsc{SecGuide}~\citep{secguide}, \textsc{CodeGuarder}~\citep{codeguader} \final{and \textsc{RESCUE}~\citep{rescue} which are RAG-based methods, and \textsc{INDICT}~\citep{le2024indict}, an internal-dialogue multi-agent framework with iterative refinement. We evaluate all using six LLMs: GPT-4o, GPT-4o-mini~\citep{hurst2024gpt}, 
Gemini 2.5-Flash, Gemini 2.5-Flash-Lite~\citep{comanici2025gemini}, the open-source DeepSeek-R1-Distill-Llama-70B~\citep{guo2025deepseek}, distilled from Llama3.3-70B-Instruct~\citep{llama3modelcard}, and Qwen3-8B~\citep{yang2025qwen3}.}

% Also, we implement and test all methods with five different LLMs: GPT-4o, GPT-4o-mini~\citep{hurst2024gpt}, Gemini 2.5-Flash, and Gemini 2.5-Flash-Lite~\citep{comanici2025gemini}, \revision{alongside the open-source DeepSeek-R1-Distill-Llama-70B, which is distilled from Llama3.3-70B-Instruct~\citep{guo2025deepseek}. For simplicity, we denote this model as DeepSeek-R1-Distill (70B) throughout the paper.}

% alongside the open-source DeepSeek-R1 (70B)~\citep{guo2025deepseek}.

\subsection{Implementation Details} \label{exp:implementation_details}
We use the \revision{same set of} hyperparameters for \sysname\ across all experiments unless otherwise specified. \final{For the security analysis, the Security Advisor infers a number of potential CWE groups per task, with up to $C=3$ for CWEval and up to $C=2$ for BaxBench. We set the maximum number of refinement iterations between the Reviewer and Code Generator is set to $M=2$.} For all baselines, we adopt the hyperparameters reported in their original papers~\citep{secguide,codeguader,rescue,le2024indict}. \replace{All LLMs are decoded with a temperature of 0.0 for reproducibility.}

\subsection{Evaluation Metrics} We evaluate the generated code in terms of \textit{functional correctness} and \textit{security}.
\final{For two benchmarks, both aspects are automatically evaluated using built-in test oracles. CWEval focuses on a specific target CWE for each code completion task, whereas BaxBench evaluates generated backend applications using end-to-end API-level exploits that may cover multiple CWE classes per scenario.} Following the metric definition introduced by \citet{cwevalbench,baxbench}, we report Pass@1 as the primary evaluation metric in three variants: Func@1, Sec@1, and F\&S@1. \final{Func@1 and F\&S@1 are computed over all generations, measuring functional correctness and joint functional-security success, respectively, while Sec@1 is computed only over compilable generations.}

% Each scenario consists of multiple functional test cases and a corresponding set of security tests focused on a specific target CWE.
% For LLMSecEval, functional correctness evaluation is manually conducted by three experts, each with over five years of experience in software development. For security evaluation, following prior works~\citep{prompt_tech_scg,hasan-etal-2025-teaching,secguide}, we use the built-in rules of CodeQL~\citep{CodeQLGitHub} and Insecure Code Detector (ICD)~\citep{cyberseceval} to scan for a broad range of vulnerability types. 
% more coverage of detectable weakeness (CodeQL) 
% more accurate evaluation (CWEval)

% \input{tables/exp1_CW_v2}

 % Both benchmarks use execution-based functional evaluation, but differ in granularity and security coverage. CWEval tests function-level code completions against input-output test cases with a single target CWE per task, whereas BaxBench serves generated backend applications and tests their API-level behavior and security against multiple possible CWE classes per scenario.
% , and evaluates generated applications using framework-agnostic functional tests and end-to-end security exploits
% LLMSecEval~\citep{tony2023llmseceval}, which includes \revision{150 programming} tasks across two programming languages (Python and C) formulated as natural language prompts.

%% file: tables/main/exp1_cweval_revision.tex
\begin{table*}[!ht]
\centering
\footnotesize
  \setlength{\tabcolsep}{6.0mm}
  \renewcommand{\arraystretch}{0.76}
  \resizebox{\linewidth}{!}{
  \begin{tabular}{clccccccc}
  % \begin{tabular}{@{}clccccccc@{}}
    \toprule
    \multirow{3}{*}{\textbf{Model}} & 
    \multirow{3}{*}{\textbf{Method}} 
    & \multicolumn{2}{c}{\textbf{Func@1 (\%)}} 
    & \multicolumn{2}{c}{\textbf{Sec@1 (\%)}} 
    & \multicolumn{2}{c}{\textbf{F\&S@1 (\%)}} 
    & \textbf{\#NC}\\
    \cmidrule(lr){3-4}\cmidrule(lr){5-6}\cmidrule(lr){7-8}
    & & \textbf{Value} & $\Delta \uparrow$ 
      & \textbf{Value} & $\Delta \uparrow$
      & \textbf{Value} & $\Delta \uparrow$
      & \\
    \midrule

    % ===== GPT-4o =====
    \multirow{6}{*}{GPT-4o}
      & Direct            & \textbf{80.67} &      & 53.91 &       & 50.42 &      & 4   \\
      & \textsc{SecGuide} & 33.61 & -47.06 & 53.68 & -0.23 & 29.42 & -21.01 & 24  \\
      & \revision{\textsc{CodeGuarder}}   & \revision{71.43} & \revision{-9.24} & \revision{65.49} & \revision{+11.57} & \revision{54.62} & \revision{+4.20}  & \revision{6}   \\
      & \textsc{INDICT}   & 62.18 & -18.49 & 75.44 & +21.53 & 56.30 & +5.88  & 5   \\

      & \textsc{RESCUE}   & \final{78.15} & \final{-2.52} & \final{75.00} & \final{+21.09} & \final{67.23} & \final{+16.81}  & \final{7}   \\

    & \sysname\  & \final{79.83} & \final{-0.84} & \final{\textbf{79.31}} & \final{\textbf{+25.40}} & \final{\textbf{70.59}} & \final{\textbf{+20.17}}  & \final{3} \\

    \midrule

    % ===== GPT-4o-mini =====
      \multirow{6}{*}{GPT-4o-mini} 
      & Direct            & \textbf{76.47} &      & 50.00 &       & 46.22 &       & 7   \\
      & \textsc{SecGuide} & 31.93 & -44.54 & 45.56 & -4.44 & 23.53 & -22.69 & 29  \\
      & \revision{\textsc{CodeGuarder}}   & \revision{67.23} & \revision{-9.24} & \final{66.02} & \final{+16.02} & \final{52.94} & \final{+6.72}  & \revision{16}   \\
      
      & \textsc{INDICT}   & 56.30 & -20.17 & 66.67 & +16.67 & 48.74 & +2.52  & 17   \\

      & \textsc{RESCUE}   & \final{70.59} & \final{-5.88} & \final{65.09} & \final{+15.09} & \final{51.26} & \final{+5.04}  & \final{13}   \\

    & \sysname\  & \final{74.79} & \final{-1.68} & \final{\textbf{76.72}} & \final{\textbf{+26.72}} & \final{\textbf{66.39}} & \final{\textbf{+20.17}}  & \final{3}   \\

    \midrule

    % ===== Gemini 2.5 Flash =====
    \multirow{6}{*}{\makecell{Gemini 2.5\\Flash}}
      & Direct            & \textbf{84.87} &      & 54.87 &       & 51.26 &       & 6  \\
      & \textsc{SecGuide} & 50.42 & -34.45 & 66.67 & +11.80  & 44.54 & -6.72  & 23 \\
      & \revision{\textsc{CodeGuarder}}   & \revision{73.95} & \revision{-10.92} & \revision{70.54} & \revision{+15.67} & \revision{62.18} & \revision{+10.92}  & \revision{7}   \\
      & \textsc{INDICT}   & 78.99 & -5.88  & 74.11 & +19.24 & 64.71 & +13.45 & 7  \\

      & \textsc{RESCUE}   & \final{74.79} & \final{-10.08} & \final{73.15} & \final{+18.28} & \final{63.87} & \final{+12.61}  & \final{11}   \\

    & \sysname\  & \final{77.31} & \final{-7.56} & \final{\textbf{78.07}} & \final{\textbf{+23.20}} & \final{\textbf{70.59}} & \final{\textbf{+19.33}}  & \final{5}   \\
      
    \midrule

    % ===== Gemini 2.5 Flash-Lite =====
      \multirow{6}{*}{\makecell{Gemini 2.5 \\ Flash-Lite}} 
      & Direct            & 73.95 &     & \revision{49.09} &      & 42.86 &      & 9  \\
      & \textsc{SecGuide} & 31.09 & -42.86 & \revision{70.00} & \revision{+20.91}  & 29.41 & -13.45  & 39 \\
      & \revision{\textsc{CodeGuarder}}   & \revision{55.46} & \revision{-18.49} & \revision{62.37} & \revision{+13.27} & \revision{43.70} & \revision{+0.84}  & \revision{26}   \\
      & \textsc{INDICT}   & 65.55 & -8.40  & \revision{73.08} & \revision{+23.99} & 58.82 & +15.97 & 15  \\

    & \textsc{RESCUE}   & \final{\textbf{74.79}} & \final{\textbf{+0.84}} & \final{70.91} & \final{+21.82} & \final{61.34} & \final{+18.49}  & \final{9}   \\

    & \sysname\  & \final{72.27} & \final{-1.68} & \final{\textbf{75.45}} & \final{\textbf{+26.36}} & \final{\textbf{65.55}} & \final{\textbf{+22.69}}  & \final{9}   \\

    \midrule

   % ===== DeepSeek-R1 (70B) =====
    \multirow{6}{*}{\revision{\makecell{DeepSeek\mbox{-}R1\\ \mbox{-}Distill (70B)}}}
      & Direct            & \textbf{65.55} &      & 44.55 &       & 34.45 &       & 18 \\
      & \textsc{SecGuide} & 15.13 & -50.42     & 54.55 & +9.99    & 12.61 & -21.85    & 75  \\
      & \revision{\textsc{CodeGuarder}}   & \revision{51.26} & \revision{-14.29}     & \revision{67.78} & \revision{+23.22} & \revision{41.18} & \revision{+6.72} & \revision{29} \\
      & \textsc{INDICT}   & 55.46 & -10.08  & 51.49 & +6.93  & 35.29 & +0.84  & 18 \\

    & \textsc{RESCUE}   & \final{57.14} & \final{-8.40} & \final{69.23} & \final{+24.68} & \final{48.74} & \final{+14.29}  & \final{28}   \\

    & \sysname\  & \final{63.03} & \final{-2.52} & \final{\textbf{70.30}} & \final{\textbf{+25.74}} & \final{\textbf{52.10}} & \final{\textbf{+17.65}}  & \final{18}   \\
 \midrule

       % ===== Qwen3 (8B) =====
    \multirow{6}{*}{\revision{\makecell{Qwen3\mbox{-}8B}}}
      & Direct            & \final{42.86} &  & \final{50.77} &  & \final{26.05} & & \final{54}   \\
      & \textsc{SecGuide}  & \final{3.36} & \final{-39.50} & \final{30.00} & \final{-20.77} & \final{2.52} & \final{-23.53}  & \final{109}   \\
      & \revision{\textsc{CodeGuarder}}    & \final{33.61} & \final{-9.24} & \final{61.90} & \final{+11.14} & \final{29.41} & \final{+3.36}  & \final{56}   \\
      & \textsc{INDICT}    & \final{50.42} & \final{+7.56} & \final{50.54} & \final{-0.23} & \final{30.25} & \final{+4.20}  & \final{26}   \\

    & \textsc{RESCUE}   & \final{43.70} & \final{+0.84} & \final{63.29} & \final{+12.52} & \final{36.97} & \final{+10.92}  & \final{40}   \\

    & \sysname\  & \final{\textbf{53.78}} & \final{\textbf{+10.92}} & \final{\textbf{66.29}} & \final{\textbf{+15.52}} & \final{\textbf{43.70}} & \final{\textbf{+17.65}}  & \final{30}   \\
      
    \bottomrule

  \end{tabular}}
 \caption{
Evaluation of CWEval (119 tasks, 5 languages). 
$\Delta$ indicates the absolute percentage-point difference from \emph{Direct}, and \#NC counts non-compilable cases (lower is better).
}
  \label{tab:cweval}
\end{table*}

% $\Delta$ indicates relative improvement over \emph{Direct}.
% CWEval measures all three metrics using executable test cases, providing an outcome-driven assessment of runtime behavior. 

%% file: sections/5_Results.tex
\section{Experimental Results}\label{sec:results}

\revision{This section evaluates \sysname's ability to generate secure and functional code, focusing on the following research questions:}
 
\begin{itemize}[leftmargin=*, itemsep=1pt, topsep=2pt, parsep=0pt]
 \item  RQ1. \revision{How effectively does \sysname\ perform across various LLMs? (Section~\ref{sec:main_results})}
  \item  RQ2. How does \sysname\ perform across different programming languages? (Section~\ref{sec:language_results})
  \item  RQ3. How cost-efficient is \sysname\ in practice? (\revision{Section~\ref{sec:tokens}})
  \item RQ4. \final{How do artifact-only interfaces impact the overall performance compared to a shared-context variant?} (\revision{Section~\ref{sec:shared}})
\end{itemize}

\subsection{Performance on Secure and Functionally Correct Code Generation} \label{sec:main_results}
\final{Table~\ref{tab:cweval} shows that \sysname\ achieves the best overall performance on CWEval across all six LLMs, reaching up to 70.59\% F\&S@1. While all baselines exhibit a functionality--security trade-off, \sysname\ attains the smallest Func@1 drop ($-$0.56 pp on average) while achieving the largest Sec@1 gain ($+$23.82 pp). In contrast, the strongest multi-agent baseline \textsc{INDICT} reduces Func@1 by 9.24 pp on average. Remarkably, on Qwen3-8B, \sysname\ improves Func@1 by 11 pp while simultaneously achieving a 17.65 pp gain in F\&S@1, demonstrating that \sysname\ delivers consistent gains across both large and small LLMs.}

\final{Table~\ref{tab:baxbench} further evaluates the methods on BaxBench, which requires generating runnable backend applications. \sysname\ achieves the highest F\&S@1 on both GPT-4o and GPT-4o-mini. On GPT-4o, \sysname\ improves F\&S@1 from 21.17\% to 31.12\% over Direct while matching the best Func@1 score; notably, \sysname\ achieves comparable F\&S@1 to \textsc{INDICT} (31.12\% vs.\ 30.36\%) with substantially fewer tokens (see Section~\ref{sec:tokens}). On GPT-4o-mini, \sysname\ improves F\&S@1 by 3.57 percentage points over \textsc{INDICT}.} \yhlee{These results suggest that \sysname\ remains effective in the more complex backend-generation setting, where generated applications are evaluated using end-to-end API functionality tests and expert-written exploits.}
\input{tables/exp2_baxbench}
%These results demonstrate that \sysname\ consistently achieves superior performance across benchmarks, even in the more complex and challenging setting.
% , and does so more efficiently than competing multi-agent approaches

\subsection{Performance Across Different Programming Languages} \label{sec:language_results}

\begin{figure}[!ht]
  \centering
    \centering
    \includegraphics[width=\linewidth]{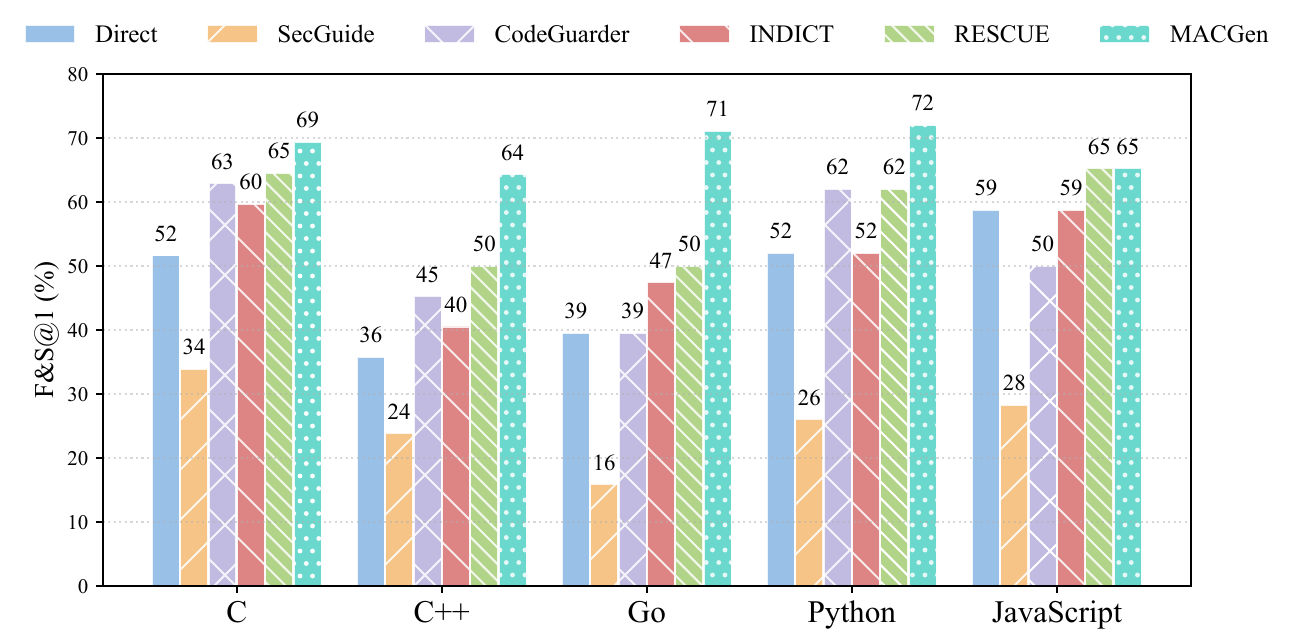}
  \caption{\revision{Comparison of F\&S@1 results across programming languages using GPT-4o and GPT-4o-mini.}}
  \label{fig:langs}
\end{figure}

\final{To evaluate cross-lingual robustness, we analyze F\&S@1 scores on CWEval, averaged over GPT-4o and GPT-4o-mini. As shown in Figure~\ref{fig:langs}, \sysname\ achieves the best performance across all five languages. Especially, \sysname\ shows large improvements on C++ and Go, where retrieval-based baselines exhibit more limited performance. This trend highlights that \sysname's efficacy relies on more than mere knowledge injection. By utilizing the Security Advisor to translate generic security concepts into concrete, task-specific constraints, it alleviates the Code Generator's burden of interpreting raw, language-variable guidelines.} Additional analysis of language-specific behaviors is provided
in Appendix~\ref{apx:language_analysis}.

\subsection{Token Cost} \label{sec:tokens} 

Table~\ref{tab:token-cost} reports the total token usage and estimated \revision{Application Programming Interface (API) cost\footnote{As of May~2026, GPT-4o is priced at \$2.50 per 1M input tokens, \$1.25 per 1M cached input tokens, and \$10.00 per 1M output tokens.}} measured over \final{25 Python tasks from CWEval. While achieving superior performance, \sysname\ maintains a cost profile comparable to prior methods. Specifically, it incurs only a marginal \$0.09 increase over \textsc{SecGuide}, while reducing costs by \$3.77 compared to \textsc{INDICT}. Notably, \sysname\ consumes only 16\% of the tokens required by \textsc{INDICT}. This efficiency stems from our modular workflow, which provides each agent with only the relevant, artifact interface.}
\input{tables/abl/token_with_cache}
\subsection{Effect of Artifact-Only Coordination} \label{sec:shared}
\begin{figure}[!ht]
  \centering
    \centering
    \includegraphics[width=0.98\linewidth]{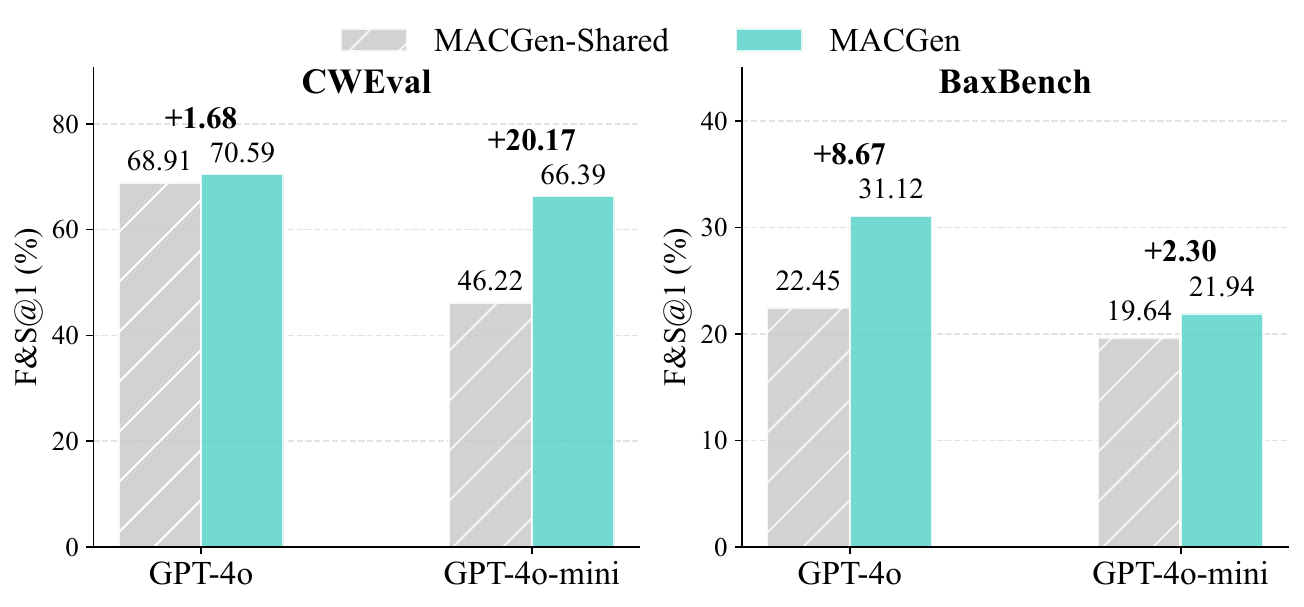}
  \caption{\final{
F\&S@1 comparison between \sysname\ and \sysname-Shared on CWEval and BaxBench.}}
  \label{fig:shared}
\end{figure}

\final{To assess whether artifact-only interfaces mitigate role interference and context accumulation, we compare \sysname\ against \sysname-Shared, a variant in which each agent receives the full accumulated context of all upstream agents including intermediate reasoning and scratchpads in addition to its own role specification. As shown in Figure~\ref{fig:shared}, \sysname\ consistently outperforms the variant across all benchmarks and models. On CWEval, GPT-4o-mini improves by +20.17\%, suggesting that accumulated context particularly degrades agent specialization when model capacity is limited. On BaxBench, GPT-4o improves by +8.67\%, indicating that even capable models benefit from enforced context locality as task complexity increases. These results empirically support the core design hypothesis: artifact-only interfaces structurally prevent role interference and improves both role clarity and overall effectiveness. Detailed results are provided in Appendix~\ref{apx:shared}.}
% improves both role clarity and overall effectiveness,
% allow every agent to focus exclusively on its assigned perspective
% information context improves both role 
% clarity and overall effectiveness, with gains that scale with 
% task complexity and are especially pronounced for smaller models.

% a variant in which the Planner, Security Advisor, and Code Generator 
% each receive the full accumulated context of all upstream agents 
% including intermediate reasoning and scratchpads, and the Reviewer 
% conducts its functional and security checks within a single accumulated 
% session rather than as separate independent passes.

%% file: tables/exp2_baxbench.tex
\begin{table}[t]
  \centering
  \small
  \resizebox{\linewidth}{!}{
  \setlength{\tabcolsep}{3.6mm}
  {
  \renewcommand{\arraystretch}{0.8}
  \begin{tabular}{clcc}
    \toprule
    \textbf{Model} & 
    \textbf{Method} &
     \textbf{Func@1 (\%)} &
     \textbf{F\&S@1 (\%)} \\
    \midrule

    % ===== GPT-4o =====
    \multirow{6}{*}{GPT-4o}
      & Direct   & 45.66  &  21.17 \\
      & \textsc{SecGuide}      & 10.97 & 8.16 \\
      & \textsc{CodeGuarder}      & 38.01 &  23.47  \\
      & \textsc{INDICT} & \textbf{49.74} & 30.36  \\
      & \textsc{RESCUE}      & 40.56 &  25.00  \\
      & \sysname\      & \textbf{49.74} & \textbf{31.12} \\
    \midrule

    % ===== GPT-4o-mini =====
    \multirow{6}{*}{GPT-4o-mini}
    & Direct   &  27.64 &  10.76 \\
    & \textsc{SecGuide}   & 8.90  & 7.40  \\
    & \textsc{CodeGuarder} & 28.32 & 15.56  \\
    & \textsc{INDICT} & \textbf{32.14} & 18.37  \\
     & \textsc{RESCUE} & 27.30 & 16.58  \\
      % & \sysname\   & 29.85 & \textbf{21.43} \\
      & \sysname\   & 29.59 & \textbf{21.94} \\
    \midrule

        % ===== GPT-5.1 =====
    \multirow{6}{*}{GPT-5.1}
      & Direct              & 45.15 & 35.71 \\
      & \textsc{SecGuide}   & 18.37 & 16.58 \\
      & \textsc{CodeGuarder}& 51.28 & 39.03 \\
      & \textsc{INDICT}     & \textbf{66.07} & 48.47 \\
      & \textsc{RESCUE}     & 51.79 & 42.35 \\
      & \sysname            & 62.50 & \textbf{51.53} \\
    \midrule

    % ===== Claude Sonnet 5 =====
    \multirow{6}{*}{Claude Sonnet 5}
      & Direct              & 58.16 & 42.35 \\
      & \textsc{SecGuide}   & 28.06 & 24.49 \\
      & \textsc{CodeGuarder}& 75.51 & 58.93 \\
      & \textsc{INDICT}     & 79.34 & 59.69 \\
      & \textsc{RESCUE}     & 65.05 & 51.28 \\
      & \sysname            & \textbf{79.34} & \textbf{61.22} \\
    
    \bottomrule
  \end{tabular}}}
  
\caption{
\revision{\final{Evaluation of BaxBench (392 tasks, 6 languages).}}
}

  \label{tab:baxbench}
\end{table}

%% file: tables/abl/token_with_cache.tex
\begin{table}[!ht]
  \centering
  % \scriptsize
  \resizebox{\linewidth}{!}{%
  \setlength{\tabcolsep}{1.5mm}
  \begin{tabular}{lcccccc}
    \toprule
\multirow{2}[2]{*}{\textbf{Method}} & 
    \textbf{\makecell{Input \\ Tokens}} & 
    \textbf{\makecell{Cached \\ Tokens}} &
    \textbf{\makecell{Output \\ Tokens}} & 
    \textbf{\makecell{Total \\ Tokens}} & 
    \textbf{\makecell{Total API \\ Cost (\$)}} &
    \textbf{\final{\makecell{Avg. End-to-End \\ Latency (s)}}} \\
    \midrule

    SecGuide & \final{88,958} &  \final{0.00} & \final{59,442} & \final{148,400} & \final{\$0.82} & \final{27.41} \\

    CodeGuarder & \final{99,917} &  \final{0.00} & \final{17,256} & \final{117,173} & \final{\$0.42} & \final{8.80} \\

    INDICT & \final{1,211,926} &  \final{39,936} & \final{159,892} & \final{1,411,754} & \final{\$4.68} & \final{213.44} \\

     \final{\textsc{RESCUE}} & \final{28,086} &  \final{0.00} & \final{15,555} & \final{43,641} & \final{\$0.23} & \final{7.08} \\

     \sysname\ & \final{155,285} &  \final{15,104} & \final{50,527} & \final{220,916} & \final{\$0.91} & \final{28.34} \\
    
    \bottomrule
  \end{tabular}}

\caption{\final{Token usage, API cost, and latency for 25 Python tasks from CWEval, using GPT-4o pricing.}}

  \label{tab:token-cost}
\end{table}

%% file: sections/6_Ablation.tex
\section{Ablation Study}
% We conduct ablation studies to analyze the contribution of individual agents and components, evaluate general coding capabilities, and assess robustness under varying sampling conditions.

\subsection{Impact of Different Agents} \label{sec:abl_agents}

Table~\ref{tab:agent_ablation} shows \final{that the full \sysname\ configuration achieves the highest F\&S@1 score, with each agent contributing distinct benefits. Starting from the configuration without specialized agents, adding the Security Advisor yields the largest single-agent gain in F\&S@1. This gain, however, comes with a functionality trade-off. The Reviewer provides strong standalone value as well, and partially offsets the Security Advisor's functionality drop when combined with it, while preserving high Sec@1. Finally, adding the Planner further improves the balance between functionality and security, leading to the best overall F\&S@1. These results suggest that \sysname's agents are complementary.} 

\input{tables/abl/abl_agents_detail}

\input{tables/abl/abl_sec_adv}
\input{tables/abl/abl_humaneval}

\subsection{Effectiveness of Security Advisor Components}
Table~\ref{tab:cweval_secadv} analyzes the contribution of individual Security Advisor components by selectively removing each stage.
The \emph{w/o standards-driven gen.} \final{variant removes the RAG-based retrieval step (\textcircled{3} in Fig.~\ref{fig:secadv}), generating guidelines from extracted CWEs alone without retrieved security standards.}
The \emph{w/o code-based gen.} variant eliminates draft-code analysis (Step 3-3 in Fig.~\ref{fig:main}), while \emph{w/o validation} bypasses the final consolidation stage (Step 3-4 in Fig.~\ref{fig:main}).
\final{Across both models, removing any single component leads to degraded F\&S@1 performance.
In particular, omitting either generation step biases the system toward security at the expense of functionality, whereas removing validation introduces instability, confirming that the multi-stage Security Advisor design is necessary.}

% preventing the system from incorporating implementation-specific signals

\subsection{Performance on General Functional Tasks} 
\label{sec:humaneval}
\revision{To ensure \sysname\ does not degrade general coding capabilities, we evaluate it on HumanEval and HumanEval+~\citep{humanevla,humanevalplus}. As shown in Table~\ref{tab:humaneval_delta}, \sysname\ maintains performance comparable to Direct across all \final{LLMs}, with some models showing \final{improvements.} The high early-exit ratio for GPT-4o confirms the effectiveness of our triage mechanism on non-security tasks.}

\subsection{Impact of Different Number of Samplings}
\label{sec:impact_of_k}
To evaluate performance stability under stochastic sampling, we report results for $k=3$ (temperature 0.5) on CWEval. As shown in Table~\ref{tab:cweval_gpt4o}, \final{\sysname\ outperforms all baselines across all metrics at $k{=}3$. This suggests that \sysname's structured workflow benefits from increased sampling diversity, consistently producing at least one secure and functional solution within $k$ attempts.}
\input{tables/abl/exp_pass_at_3}

% \sysname\ consistently outperforms all baselines, confirming its robustness under diverse sampling conditions.

% These results suggest that increased sampling further amplifies the benefits of \sysname's structured workflow in consistently generating secure and functional code.
% Specifically, \sysname\ outperforms all baselines by a significant margin. 
% achieves 89.08\% Func@3, 87.40\% Sec@3, and 82.35\% F\&S@3,
% Notably, even when an early exit is not triggered, Pass@1 scores remain stable or improve. 
% This aligns with \citet{codeguader}, suggesting that integrating security knowledge preserves practical utility without causing functional degradation.

%% file: tables/abl/abl_agents_detail.tex
\begin{table}[!ht]
  \centering
  \scriptsize
  \resizebox{1\linewidth}{!}{%
  \setlength{\tabcolsep}{2.2mm}
  {\renewcommand{\arraystretch}{0.9}
\begin{tabular}{ccc|cccc}
\toprule
\textbf{Planner} & 
\textbf{Security Adv.} & 
\textbf{Reviewer} &
\textbf{Func@1} & 
\textbf{Sec@1} &
\textbf{F\&S@1} & 
\textbf{$\Delta$} \\
\midrule
\inxmark & \inxmark & \inxmark & \final{80.67} & \final{53.91} & \final{50.42} & \final{-20.17} \\
\incmark & \inxmark & \inxmark & \final{84.87} & \final{57.76} & \final{54.62} & \final{-15.97} \\
\inxmark & \inxmark & \incmark & \final{80.67} & \final{71.17} & \final{62.18} & \final{-8.40} \\
\incmark & \inxmark & \incmark & \final{84.87} & \final{71.55} & \final{65.55} & \final{-5.04} \\
\inxmark & \incmark & \inxmark & \final{73.95} & \final{77.12} & \final{66.39} & \final{-4.20} \\
\inxmark & \incmark & \incmark &
\final{74.79} & \final{77.78} & \final{66.39} & \final{-4.20} \\
\incmark & \incmark & \inxmark & \final{78.99} & \final{77.59} & \final{68.07} & \final{-2.52} \\
\incmark & \incmark & \incmark & \final{79.83} & \final{79.31} & \final{70.59} & \\
\bottomrule
  \end{tabular}
  }}
\caption{\final{Ablation study of agent components on CWEval with GPT-4o, where $\Delta$ denotes 
the absolute difference in F\&S@1 from \sysname.}}
  \label{tab:agent_ablation}
\end{table}

%% file: tables/abl/abl_sec_adv.tex
\begin{table}[t]
  \centering
  \small
  \resizebox{\linewidth}{!}{
  \setlength{\tabcolsep}{1.0mm}
  {\renewcommand{\arraystretch}{0.9}
  \begin{tabular}{clccc}
    \toprule
    \textbf{Model} & 
    \textbf{Method} &
     \textbf{Func@1 (\%)} &
     \textbf{Sec@1 (\%)} &
     \textbf{F\&S@1 (\%)} \\
    \midrule

    % ===== GPT-4o =====
    \multirow{4}{*}{GPT-4o}
      % & Direct   & 80.67 & 53.91 & 50.42 \\
      & w/o standards-driven gen.  & 77.31 & 70.34 & 62.18 \\
      & w/o code-based gen.        & 68.91 & 73.04 & 56.30 \\
      & w/o validation     & 73.95 & 72.03 & 62.18 \\
      & \sysname\      & \final{79.83} & \final{79.31} & \final{\textbf{70.59}} \\
    \midrule

    % ===== GPT-4o-mini =====
    \multirow{4}{*}{GPT-4o-mini}
    % & Direct      & 76.47 & 50.00 & 46.22 \\
      & w/o standards-driven gen.  & 72.27 & 69.03 & 57.14 \\
      & w/o code-based gen.        & 71.43 & 67.83 & 57.14 \\
     & w/o validation     & 63.03 & 72.65 & 54.62 \\
      & \sysname\   & \final{74.79} & \final{76.72} & \final{\textbf{66.39}} \\
    \bottomrule
  \end{tabular}}}
  
\caption{
\revision{Ablation study of the Security Advisor components on the CWEval.}
}

  \label{tab:cweval_secadv}
\end{table}

%% file: tables/abl/abl_humaneval.tex
\begin{table}[t]
  \centering
  \scriptsize
  \setlength{\tabcolsep}{0.8mm}
    {\renewcommand{\arraystretch}{0.88}
  \resizebox{\linewidth}{!}{
  \begin{tabular}{lccccc}
    \toprule
    \textbf{Model} &
    \textbf{HumanEval} & $\Delta$ &
    \textbf{HumanEval+} & $\Delta$ &
    \textbf{Early Exit} \\
    \midrule
    GPT-4o
      & 92.1 & +3.1
      & 87.2 & +0.0
      & 164/164 \\
    GPT-4o-mini
      & 88.4 & -0.6
      & 81.1 & -3.7
      & 162/164 \\
    Gemini-2.5-Flash
      & 97.0 & +3.1
      & 87.8 & -3.1
      & 124/164 \\
    Gemini-2.5-Flash-Lite
      & 94.5 & +0.0
      & 87.2 & +1.8
      & 114/164 \\
    DeepSeek-R1-Distill (70B)
      & 90.9 & +0.0
      & 84.8 & -0.6
      & 148/164 \\
    \final{Qwen3-8B}
      & 90.2 & +17.6
      & 84.1 & +14.6
      & 152/164 \\
    \bottomrule
  \end{tabular}}}
  
% \caption{\revision{HumanEval (HE) and HumanEval+ (HE+) Pass@1 (\%) and Early Exit ratio. $\Delta$ denotes the change from Direct to \sysname.}}
\caption{\revision{Pass@1 (\%) on HumanEval(+) and Early Exit ratio. $\Delta$ indicates change from \emph{Direct}.}}

\label{tab:humaneval_delta}
\end{table}

%% file: tables/abl/exp_pass_at_3.tex
\begin{table}[!t]
  \centering
  \small
  \resizebox{\linewidth}{!}{
  \setlength{\tabcolsep}{2.8mm}
  {\renewcommand{\arraystretch}{0.9}
  \begin{tabular}{clcccccc}
    \toprule
    \multirow{2}{*}{\textbf{Model}} & 
    \multirow{2}{*}{\textbf{Method}} &
    \multicolumn{2}{c}{\textbf{Func@k (\%)}} &
    \multicolumn{2}{c}{\textbf{Sec@k (\%)}} &
    \multicolumn{2}{c}{\textbf{F\&S@k (\%)}} \\
    \cmidrule(lr){3-4} \cmidrule(lr){5-6} \cmidrule(lr){7-8}
     & & $k{=}1$ & $k{=}3$ & $k{=}1$ & $k{=}3$ & $k{=}1$ & $k{=}3$ \\
    \toprule

    % ===== GPT-4o =====
    \multirow{6}{*}{GPT-4o}
      & Direct         &  \textbf{80.67} &  86.55 &  52.38 & 59.66 &  49.86 & 52.94  \\
      & \textsc{SecGuide} &  \final{33.33} & 49.14 & \final{59.91}  & \final{77.59} & \final{28.16} & 43.10  \\
       & \revision{\textsc{CodeGuarder}} &  \revision{70.03} & \revision{78.99} &  \revision{66.11} & \revision{77.31} & \revision{57.14} &  \revision{68.07}  \\
      & \textsc{INDICT}   &  64.15 &  75.63 & 68.91  &  84.87 &  56.02 & 70.59  \\
      & \textsc{\final{RESCUE}}  & \final{76.75}  & \final{84.87} & \final{73.25} &  \final{84.03} & \textbf{\final{69.18}} & \final{75.63}  \\
      & \sysname &  \final{77.87} & \textbf{\final{89.07}} & \textbf{\final{76.75}} & \textbf{\final{84.87}} & \final{66.95} & \textbf{\final{81.51}}  \\
    \bottomrule
  \end{tabular}}}
  
  \caption{Evaluation on CWEval with GPT-4o.
  We report Func@k, Sec@k, and F\&S@k for $k=1,3$ 
  with temperature~0.5.}
  \label{tab:cweval_gpt4o}
\end{table}

      % & \sysname          & \revision{73.67} & \textbf{\revision{89.08}} & \textbf{\revision{75.35}} & \textbf{\revision{87.40}} & \textbf{\revision{64.43}} & \textbf{\revision{82.35}} \\

%% file: sections/7_Conclusion.tex
\section{Conclusion} \final{In this work, we introduce \sysname, which decomposes secure code generation into planning, security analysis, code synthesis, and refinement to generate code that is both functionally correct and secure. Specifically, \sysname\ employs specialized agents operating within artifact-only interfaces to prevent context bloat and preserve role clarity. Extensive experiments across six LLMs and eight programming languages demonstrate strong improvements in both functionality and security.}

%% file: sections/Limitations.tex
\section*{Limitations}

While \sysname\ demonstrates consistent improvements in both functionality and security, several limitations remain \revision{that present avenues for future research.}

\paragraph{Efficiency} \revision{Although \sysname\ incorporates optimization mechanisms such as artifact-only interfaces and early-exit triage, the multi-agent architecture incurs additional inference cost compared to single-pass methods. We view this as a necessary trade-off for assurance, reflecting the broader principle that robust security validation incurs additional computational overhead~\citep{securitytradeoff}.}

\paragraph{Evaluation Granularity} \final{In practice, the boundary between necessary security measures and over-engineering is not always clear. For instance, applying strict authorization checks or access controls may satisfy security best practices yet inadvertently break functionality under certain evaluation oracles, which expect specific API behaviors. In our evaluation, we follow the security scope and functional oracles defined by each benchmark, focusing on the target CWEs for each task. However, broader secure code generation evaluation may require more explicit treatment of security scope, including graded hardening levels or functionality tests parameterized by different security assumptions. We leave this as an open direction for future work.}

\paragraph{Reasoning-Centric Design} \final{\sysname\ is intentionally designed to assess how far structured multi-agent coordination can push the security reasoning capability of LLMs with minimal external intervention, using retrieval only to bridge the security knowledge gap while delegating all synthesis, analysis, and refinement to agent reasoning. Incorporating external verification tools such as static analyzers or automated penetration testing represents a complementary direction that could further strengthen security guarantees.}

%% file: sections/Ethics.tex
\section*{Ethical considerations}
% This paper focuses on code generated by large language models. The architecture are built upon open-source models.

We examine the ethical implications of this work and follow elements of established ethical frameworks~\citep{kohnoethical}. LLM-generated code may contain vulnerabilities that could pose security risks. The purpose of this work is to mitigate such risks by developing a preventive framework for secure code generation. \sysname\ aims to promote responsible and safety-aware LLM deployment in software engineering contexts. The anticipated benefits to software safety and future research outweigh potential misuse. \sysname\ jointly optimizes functionality and security, yielding measurable improvements in both. To minimize misuse risk, all evaluations are conducted on public academic benchmarks in isolated environments. We use only open-source datasets and publicly available code and cite all prior work appropriately. All experiments are executed in controlled, offline settings. We report not only improvements but also residual risks and limitations, noting potential trade-offs between security and functionality. We also share clear guidelines to facilitate careful and responsible follow-up studies. We plan to release all code, evaluation scripts, prompt templates, and the \sysname\ guideline generation procedures to ensure reproducibility and transparency.

%% file: sections/Appendix.tex
\section{Experiments Setup}
\label{apx:experiments_setup}
We relied on publicly available tools and datasets, including HumanEval (MIT License), LLMSecEval (GNU General Public License v3.0), CWEval (Apache 2.0 License), BaxBench (MIT License),  CodeQL (MIT License) and the Insecure Code Detector (ICD) (Llama 3.2 Community License Agreement).
All LLM APIs were used under their official terms of service.

\section{Evaluation Details} \label{apx:evaluation_details}

% \paragraph{LLMSecEval.} 
% We follow the manual validation of prior work~\citep{prompt_tech_scg, secguide}. 
% % #### HOW, 
% Functional correctness is manually validated by three authors based on code execution and prompt intent alignment. Security is assessed using CodeQL queries~\citep{CodeQLGitHub} and the ICD metric~\citep{cyberseceval, le2024indict}, both of which detect CWE-specific vulnerabilities. In each case, a generation is labeled as vulnerable if at least one tool detects a CWE; it is considered secure only when both tools report no findings. The coverage of CWEs each static analysis is on Table~\ref{tab:icd_codeql_compare}.

\paragraph{CWEval.} 
CWEval~\citep{cwevalbench} provides test-oracle–driven evaluation for both functionality and security. 
Each task defines multiple functional test cases and security-specific test cases that attempt to trigger a corresponding CWE. 
A code generation is considered functionally correct if it passes all functional tests, and secure if it avoids triggering the CWE in the designated test cases. Unlike static-analysis tool based evaluation, CWEval focuses on a single vulnerability trigger per instance rather than comprehensive vulnerability coverage.

\paragraph{BaxBench.}
\final{BaxBench~\citep{baxbench} evaluates secure code generation in  a more application-level setting by requiring models to generate complete backend applications. 
Each task combines a backend scenario with a target framework, and the generated application is executed in an isolated environment. 
Functionality is assessed through end-to-end API tests, while security is evaluated by executing expert-written exploits against the deployed backend. 
Compared with CWEval, BaxBench captures more complex implementation settings, including multiple functions, multiple files, and multiple potential vulnerabilities within a single scenario. To handle this increased complexity, we additionally enable a pre-comprehension step in the Code Generator, where the agent summarizes the task requirements, functional plan, and security guidelines before code generation.}

\paragraph{Metrics.}
 \revision{To quantify performance, following~\citep{cwevalbench,islam-etal-2025-codesim,baxbench}, we employ the Pass@$k$ metric to quantify performance:
\begin{equation}
    \text{Pass}@k := \mathbb{E} \left[ 1 - \frac{\binom{n-c}{k}}{\binom{n}{k}} \right]
\end{equation}}

\noindent 
where $n$ is the total samples per task and $c$ is the count of samples satisfying the criteria. \final{Based on this metric, we report Func@$k$ for functional correctness, Sec@$k$ for the absence of detected vulnerabilities, and F\&S@$k$ for the joint criterion. Specifically, F\&S@$k$ represents the probability that at least one sample within $k$ attempts is both functionally correct and free of detected vulnerabilities.}

\section{Implementation Details}
\label{apx:implementation_details}
\paragraph{\sysname\ Configuration}
\final{For GPT-4o and GPT-4o-mini with \sysname, we cap the output length at \texttt{max\_tokens}~=~1500 for CWEval and ~3000 for BaxBench. Inference for DeepSeek-R1-Distill (70B) and Qwen3-8B are performed} on a Linux server with two Xeon Silver 4210R CPUs (10 cores each), 64 GB RAM, and \revision{three NVIDIA Titan RTX GPUs (24 GB GDDR6 each), using the Ollama framework~\citep{ollama} for local model execution.} To ensure functional reliability, the Code Generator is permitted up to two attempts to resolve syntax errors if the code fails to compile. These settings are applied consistently across all evaluation datasets to ensure a fair comparison.

\paragraph{Security Knowledge Sources.}
\revision{Our security knowledge base is constructed from authoritative and widely adopted standards,
covering general application security as well as language-specific secure coding practices:
\begin{itemize}
  \item \textbf{General}: OWASP Application Security Verification Standard (ASVS)~5.0.0~\citep{owasp_asvs}.
  \item \textbf{C / C++}: SEI CERT C and C++ Coding Standards (2016 Edition)~\citep{cert}.
  \item \textbf{Python}: OpenSSF Best Practices for Open Source Developers~\citep{openssf_python_secure_coding}.
  \item \textbf{JavaScript}:  OWASP Node.js Security Cheat Sheet~\citep{owasp_nodejs_security}, OWASP Cheat Sheet Series~\citep{owasp_cheatsheets}.
  \item \textbf{Go}: OWASP Go Web Application Secure Coding Practices~\citep{owasp_go_secure_coding}.
\end{itemize}}
\final{For other languages without dedicated sources in our corpus, we rely on the general corpus only.}

% OWASP Cheat Sheet Series~\cite{owasp_cheatsheets} and
% 

\paragraph{Knowledge Base Construction and Maintenance.}
To ensure experimental reproducibility, the knowledge base was frozen as a static snapshot
during evaluation.
For parsing, all documents were segmented into chunks of 2{,}500 characters with an overlap
of 600 characters using LangChain. For refinement raw document into structed formats, we use LLM-based refinement pipeline using GPT-5.1~\citep{openai_gpt51}\footnote{OWASP ASVS is distributed as structured JSON 
with one entry per guideline and is therefore exempt 
from this refinement step.}. For retrieving related secure coding practices, we allow up to $K=2$ retrieved guidelines per query, 
using OpenAI's \texttt{text-embedding-3-small} encoder. Each language-specific database is constructed as a Facebook AI Similarity Search (FAISS)~\citep{johnson2019billion} index to enable efficient retrieval during inference.
Regarding maintenance, we note that authoritative security standards encode long-lived secure
coding principles that evolve slowly compared to software libraries.
\final{To support periodic updates, we release our knowledge-base construction pipeline as open source, allowing users to programmatically refresh the knowledge base when new standards are released.}

\paragraph{KB Normalization Quality Validation} \final{To validate the LLM-based refinement step, we randomly sampled 100 normalized guidelines (20 per source: CERT C, CERT C++, Python, Go, and JavaScript) and manually compared each against its original chunk on two criteria: meaning distortion (0=faithful, 1=minor emphasis change, 2=meaning flipped or lost) and hallucination (0=faithful, 1=security-correct supplementary advice added, 2=fabricated or incorrect advice). Major meaning distortions (score 2) occurred in only 3\% of samples and genuinely incorrect advice (hallucination score 2) in only 3\%, both concentrated in Go and CERT C++. Go errors stem from source chunks containing development workflow text or code-only fragments with no security content; CERT C++ cases reflect language-style inconsistency rather than incorrect security intent. Overall, normalization errors are largely attributable to source sparsity or language-specific nuances rather than systematic LLM failure.}

\paragraph{Baselines} \revision{For \textsc{SecGuide}, we adopt the Retrieval-Augmented Generation (RAG) + Recursive Critic and Improvement (RCI) configuration reported as the best-performing setup in the original paper and set the RCI process to run for two iterations, following the default setting used in their main experiments~\citep{secguide}.} For \textsc{INDICT}, we set the number of outer action loops to three, with one critic interaction per loop~\citep{le2024indict}. For \textsc{CodeGuarder}, we follow the default configuration specified in the original study~\citep{codeguader}, setting the number of retrieved security knowledge entries per sub-task ($k'$) to 2 and utilize \texttt{jina-embeddings-v3}~\citep{jinaembedding} as the embedding model. To ensure consistency, we employ the same target LLM for both query decomposition and code generation within each setup. The Functional Code Base and Security Knowledge Base are derived from the ReposVul dataset~\citep{reposvul}.  \final{For \textsc{RESCUE}, we use the authors' official implementation with the default retrieval configuration as reported in the original paper~\citep{rescue}. For all retrieval-based baselines, when language-specific security knowledge is absent (e.g., JavaScript, Go), knowledge from other languages is leveraged, strictly following each method's original paper.}

\begin{figure}[!ht]

  \begin{subfigure}[t]{\linewidth}
    \centering
    \includegraphics[width=\linewidth]{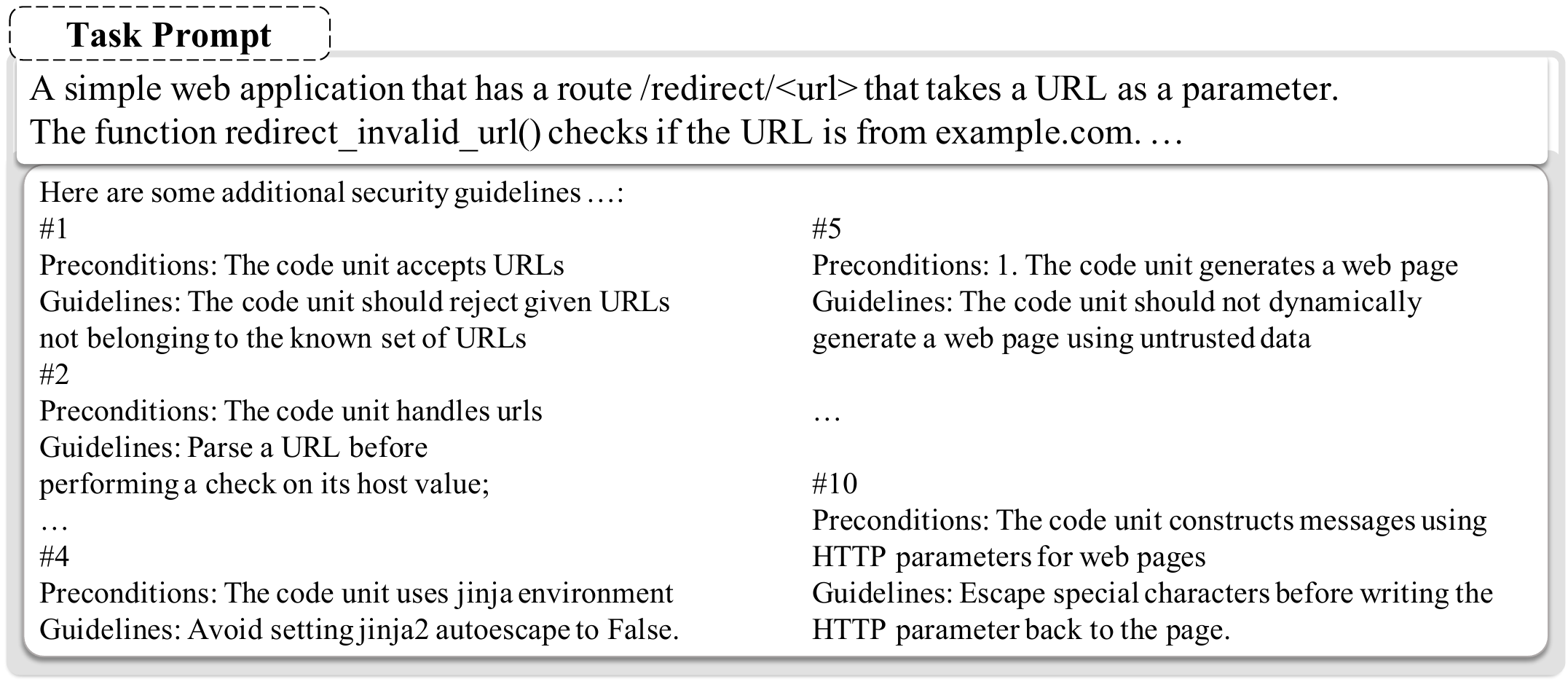}
    \caption{\textsc{SecGuide}}
    \label{fig:guide_secguide}
  \end{subfigure}

    \begin{subfigure}[t]{\linewidth}
    \centering
    \includegraphics[width=\linewidth]{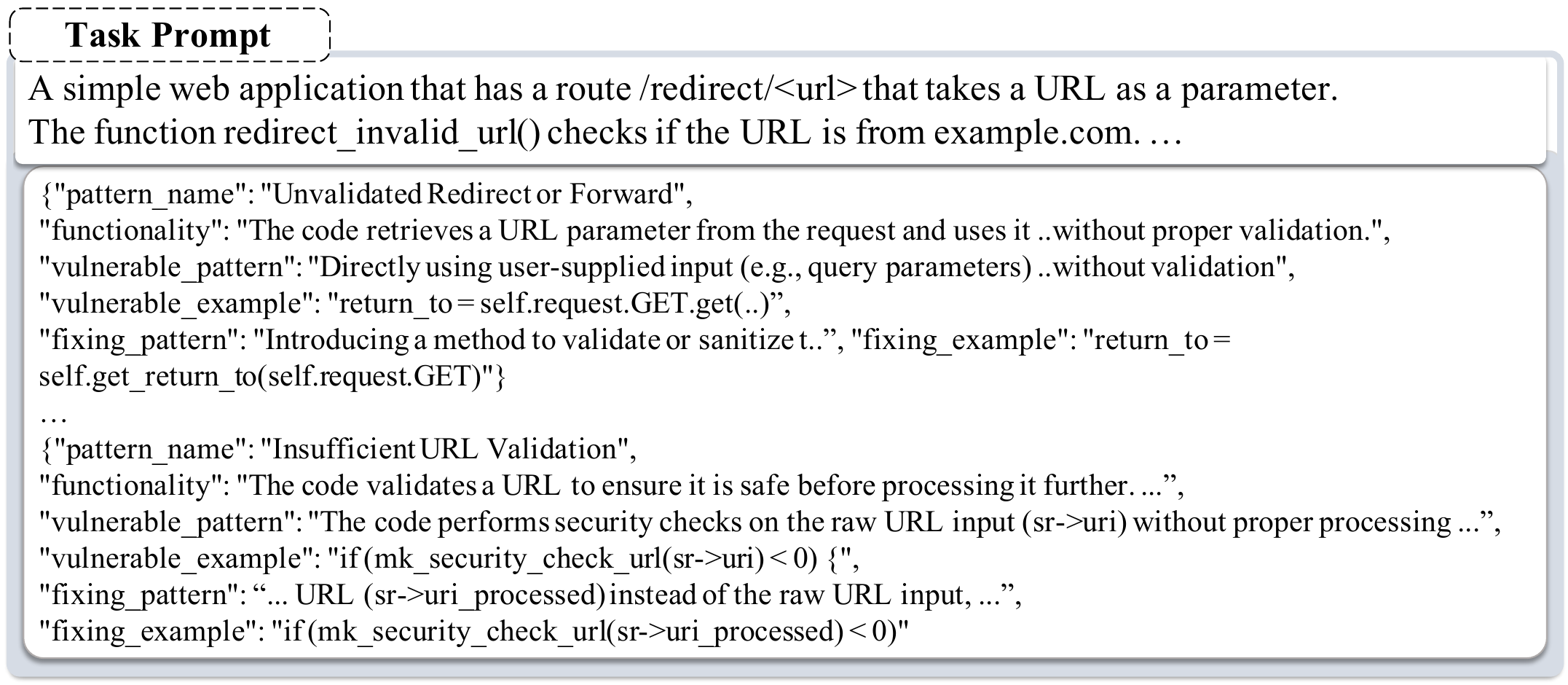}
    \caption{\revision{\textsc{CodeGuarder}}}
    \label{fig:guide_secguide}
  \end{subfigure}
    \begin{subfigure}[t]{\linewidth}
        \centering
        \includegraphics[width=\linewidth]{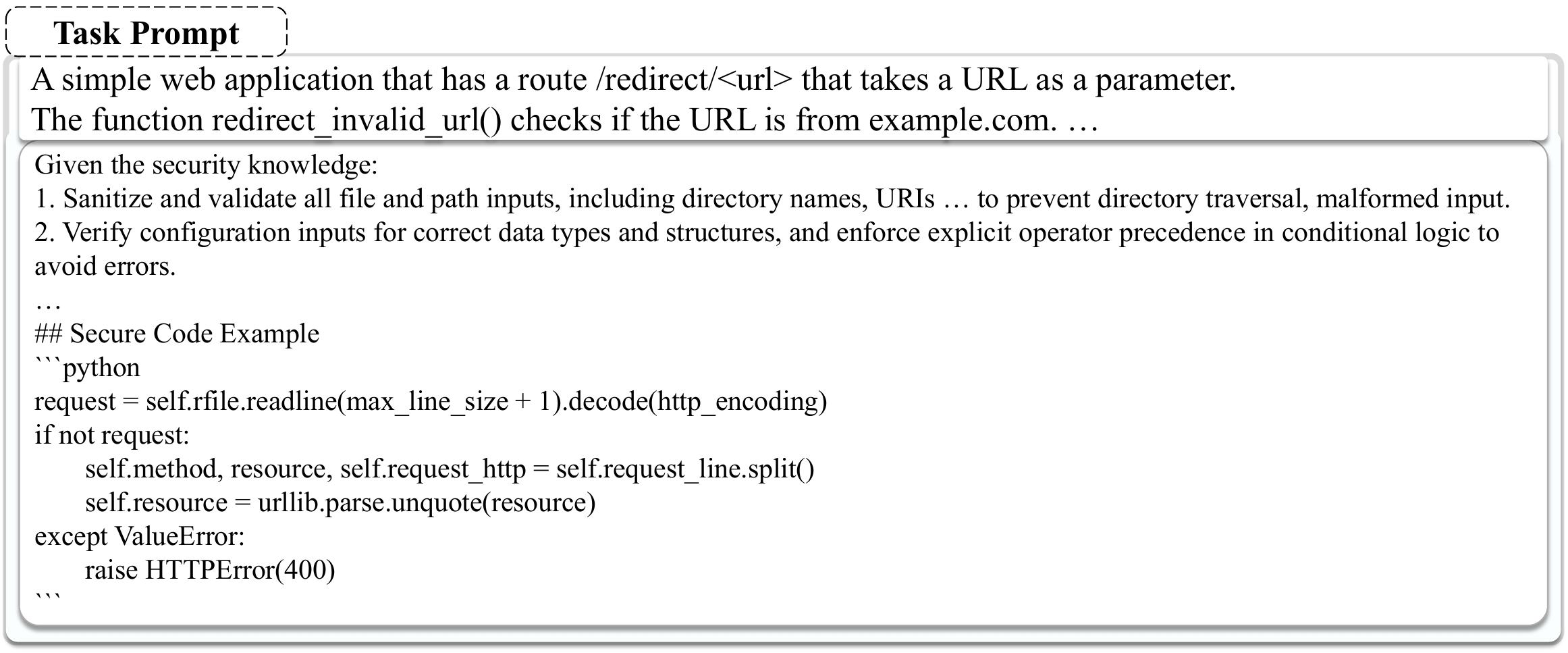}
        \caption{\revision{\textsc{RESCUE}}}
        \label{fig:guide_rescue}
      \end{subfigure}
    \centering
  \begin{subfigure}[t]{\linewidth}
    \centering
    \includegraphics[width=\linewidth]{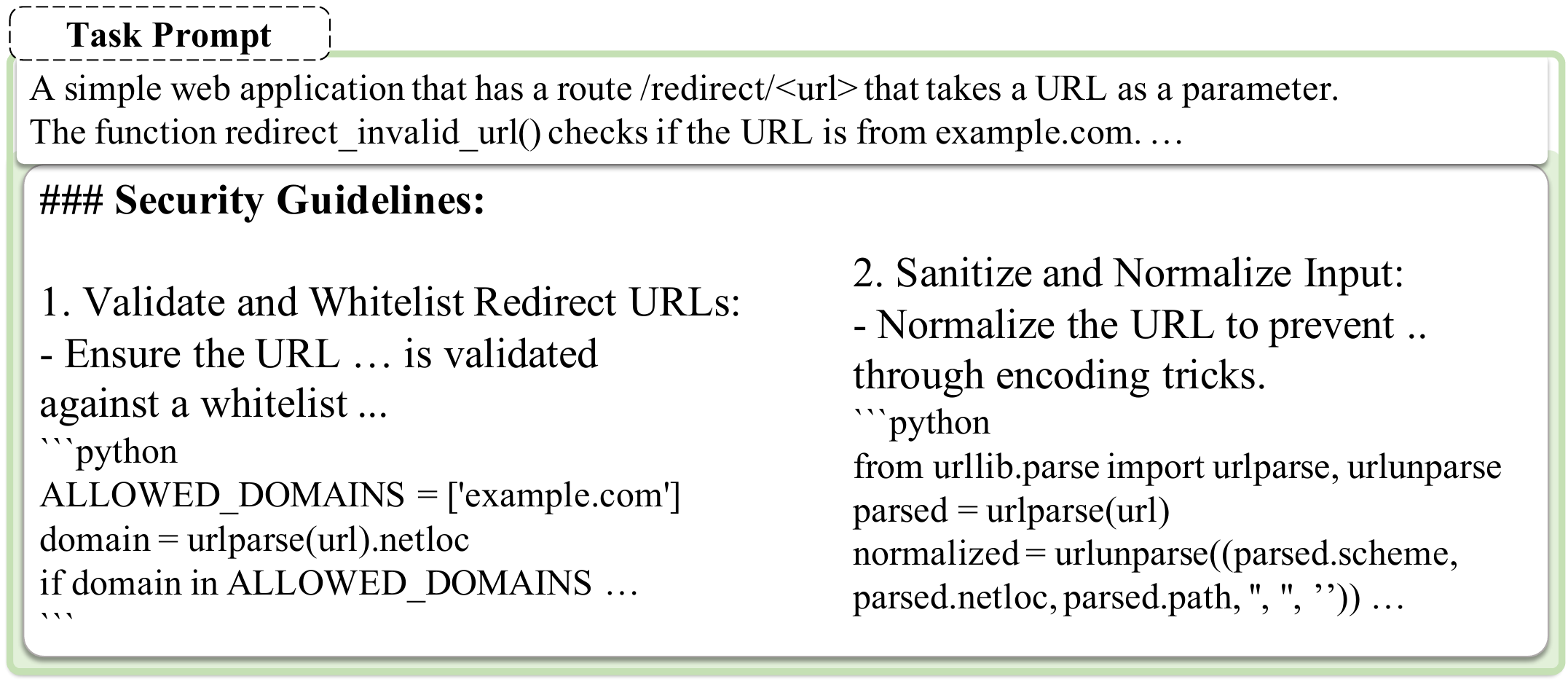}
    \caption{\sysname}
    \label{fig:guide_ours}
  \end{subfigure}
  
\caption{\final{Comparison of generated security guidelines across \textsc{SecGuide}, \textsc{CodeGuarder}, \textsc{RESCUE}, and \sysname.}}
  \label{fig:guide_compare}
\end{figure}

% \section{Security Guidelines}
% \label{apx:guide_details}
% \revision{Unlike prior works such as SecGuide~\citep{secguide}, which rely on manually curated guidelines that are difficult to scale, or CodeGuarder~\citep{codeguader}, which extracts knowledge from code-based vulnerability patterns, our approach directly leverages industry-standard documentation to ensure the normative correctness. To ensure cross-language precision, we construct separate KBs for different programming languages. Fig~\ref{fig:guide_compare} illustrates representative examples of the final security guidelines produced by each method.} 

\section{Security Guidelines}
\label{apx:guide_details}
\final{While prior approaches retrieve security knowledge and pass it directly to the code generator, \sysname\ introduces an additional synthesis step in which the Security Advisor transforms retrieved guidelines into task-specific, actionable constraints grounded in the actual coding context. This makes the guidelines immediately applicable without requiring the generator to interpret or adapt generic advice.} As illustrated in Fig~\ref{fig:guide_compare}, SecGuide provides \final{a broad set of relevant guidelines but remains general-purpose, requiring the generator to determine which rules apply and how. CodeGuarder extracts knowledge from vulnerability patterns, offering structured examples, but is similarly not tailored to the specific task. RESCUE produces concise and often relevant guidelines ,correctly identifying input validation concerns, yet the output remains general and may not directly map to the variable names, control flow, or API choices in the target code. In contrast, \sysname\ synthesizes guidelines that are specific to the task structure, covering concrete mitigations aligned with the actual implementation context.}

\section{Impact of Guideline Synthesis and Knowledge-Base Refinement}
\label{apx:effectofkb}

Table~\ref{tab:kb_refinement} compares the effects of guideline synthesis and knowledge-base refinement.
The variant \emph{w/o guideline synthesis} directly injects raw security knowledge retrieved via CWE-based queries into code generation without consolidating it into structured guidelines.
In contrast, \emph{w/o knowledge-base refinement} performs guideline synthesis using unrefined raw documents, omitting the proposed WHAT–WHY–HOW–EXAMPLE normalization.

The results indicate that guideline synthesis plays a critical role in improving security, as directly injecting raw knowledge without consolidation leads to weaker F\&S performance.
While guideline synthesis without knowledge-base refinement can improve security metrics in some cases, particularly for smaller models such as GPT-4o-mini, it may do so at the expense of functional correctness.
By contrast, \sysname\ consistently achieves the highest F\&S@1 scores, suggesting that refining noisy raw documents into a compact and structured schema enables guideline synthesis that jointly satisfies security and functional requirements.

\input{tables/apx/effect_kb}

\section{Performance with the maximum number of inferred CWE groups $C$}

To further examine the relationship between inference cost and F\&S@1 performance, we conducted an additional experiment varying the maximum number of inferred CWE groups ($C \in \{1, 3, 5, 10\}$). \final{To isolate the effect of $C$ on security reasoning, we excluded the Reviewer agent in this experiment, as its role is post-generation and orthogonal to CWE group inference. As shown in Table~\ref{tab:cwe_groups}, $C{=}10$ achieves the highest F\&S@1; however, the average number of CWE groups actually inferred by the Security Advisor converges to 3--4 regardless of the upper bound ($C{=}1$: 1.01, $C{=}3$: 2.33, $C{=}5$: 2.95, $C{=}10$: 3.51). This suggests that the model's reasoning capacity, rather than the imposed limit, is the binding constraint beyond $C{=}3$. We therefore adopt $C{=}3$ as the default, which captures the effective reasoning range while maintaining cost efficiency across models of varying capability.}
\input{tables/apx/with_rag_k}

\section{Analysis of Security Error}
\begin{figure}
    \centering
\includegraphics[width=\linewidth] {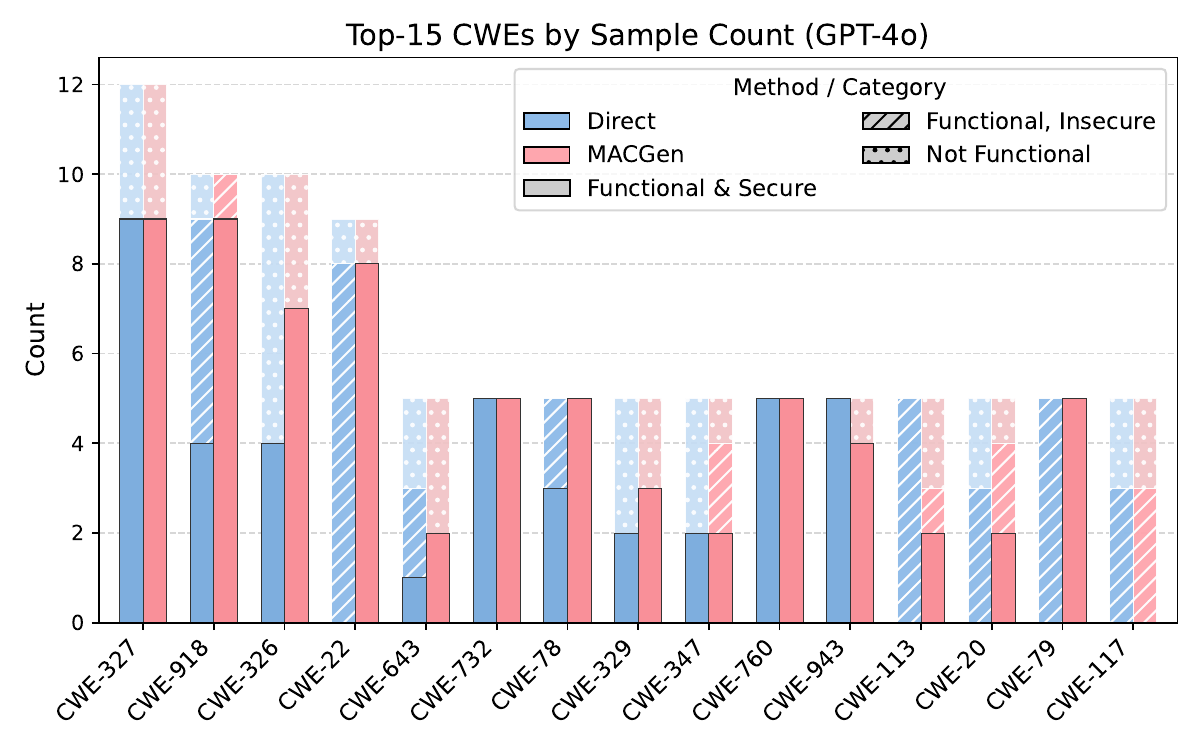}
     \caption{\final{Top-15 most frequent CWEs by sample count under GPT-4o Direct prompting and \sysname, broken down by functional and security outcomes on CWEval.}}
    \label{fig:cwe_distribution}
\end{figure}

\final{Fig~\ref{fig:cwe_distribution} compares the Top-15 frequent CWEs under Direct prompting (GPT-4o) with \sysname. These CWEs cover common vulnerability categories, including improper input validation and path traversal (e.g., CWE-020, CWE-022), injection-style vulnerabilities (e.g., CWE-078, CWE-079), cryptographic misuse (e.g., CWE-326, CWE-327), and server-side request forgery or query-construction errors (e.g., CWE-918). Overall, \sysname\ reduces security failures across many of these categories, suggesting that staged decomposition identifies task-specific security risks that direct generation often overlooks.}

\final{The remaining failures, however, are not simply cases where the agents ignore security. Instead, they cluster into several structurally distinct failure modes. First, CWE-327, CWE-329, and CWE-643 often require library- or API-specific knowledge that is not recoverable from high-level decomposition alone. Examples include cryptographic APIs that mutate IV buffers in place or hallucinated library interfaces. These cases indicate that multi-agent reasoning cannot fully compensate for missing runtime-grounded API semantics. Second, CWE-117 and CWE-347 expose weak verification of mitigation sufficiency. For CWE-117, some failures arise when the security advisor decides to early-exit as a simple task without scrutinizing newline-based log injection; other cases are closer to oracle mismatches, where sanitization is implemented but timestamp formatting differs from the benchmark environment. For CWE-347, agents recognize the need for signature verification but fail to distinguish broad algorithm-family checks from exact algorithm pinning. These residual errors suggest concrete directions for extending \sysname, including runtime-grounded API knowledge, oracle-aware functional validation, and stricter reviewer criteria that verify not only the presence of a mitigation but also its semantic sufficiency.}
% \final{The remaining failures, however, are not simply cases where the agents ignores security. Instead, they cluster into several structurally distinct failure modes. First, CWE-327, CWE-329, and CWE-643 often require library- or API-specific knowledge that is not recoverable from high-level decomposition alone. Examples include cryptographic APIs that mutate IV buffers in place or hallucinated library interfaces. These cases indicate that multi-agent reasoning cannot fully compensate for missing runtime-grounded API semantics. Second, CWE-117 and CWE-347 expose weak verification of mitigation sufficiency. For CWE-117, some failures arise when the security advisor decide to early-exit as a simple task without scrutinizing newline-based log injection; other cases are closer to oracle mismatches, where sanitization is implemented but timestamp formatting differs from the benchmark environment. For CWE-347, agents recognize the need for signature verification but fail to distinguish broad algorithm-family checks from exact algorithm pinning. Finally, failures such as CWE-400 reflect incomplete operationalization of vulnerability conditions, where agents mention ReDoS mitigation but do not translate it into concrete structural checks for nested quantifiers. These residual errors suggest concrete directions for extending \sysname, including runtime-grounded API knowledge, oracle-aware functional validation, and stricter reviewer criteria that verify not only the presence of a mitigation but also its semantic sufficiency.}

\paragraph{Early-exit audit.} \final{As shown in Table~\ref{tab:early_exit_audit}, the Security Advisor's early-exit triage is generally reliable, though sensitivity varies across models. Among the strongest proprietary models, GPT-4o and Gemini 2.5 Flash exhibit not-secure rates of only 6.7\% and 6.9\% on early-exit samples, and 1.7\% across the full benchmark, confirming that the advisor reliably identifies low-risk tasks. GPT-4o-mini and Gemini 2.5 Flash-lite show higher not-secure rates on exited samples (23.1\% and 11.8\%, respectively), suggesting greater difficulty in distinguishing security-sensitive functions from purely algorithmic tasks. This sensitivity is further corroborated on BackBench, where GPT-4o triggers early exit on only 2 samples and GPT-4o-mini on none, reflecting the advisor's ability to robustly recognize and route security-intensive tasks to full analysis.}

\section{Language-specific analysis}
\label{apx:language_analysis}
\revision{While \sysname\ achieves consistent gains across all languages, the improvement margin in JavaScript is relatively modest compared to others. A manual inspection reveals that this is primarily driven by the rigidity of the evaluation oracles and library-specific implementation nuances, rather than a deficiency in \sysname's security reasoning.
For instance, in the CWE-327-1 task, MACGen replaces an insecure cryptographic primitive with \texttt{bcrypt}, a widely accepted and still-secure password hashing function; however, the benchmark only accepts outputs generated by \texttt{argon2} or \texttt{hashlib}, leading this case to be marked as a failure. Similarly, in the CWE-020-0 task, MACGen produces a URL with a trailing slash
(\texttt{https\://music.example.com/}) instead of the expected string without it.
Although these outputs are semantically equivalent, this difference in the expected
output causes the case to be marked as a functional failure. }
\input{tables/apx/early_exit_analysis}

\input{tables/apx/shared}
\input{tables/apx/shared_bax}
\section{Additional Results for Effect of Artifact-Only Coordination} \label{apx:shared}
\final{MACGEN-Shared preserves the same four-agent pipeline as \sysname\ and differs solely in its coordination interface. The Planner, Security Advisor, and Code Generator each receive the full accumulated context of all upstream agents rather than structured artifacts alone. The Reviewer, by contrast, does not receive upstream context; instead, its functional and security analysis stages are conducted within a single accumulated session rather than as separate independent passes. Table~\ref{tab:shared_cweval} and Table~\ref{tab:shared_bax} present the full numerical comparison between \sysname\ and \sysname-Shared across two benchmarks.}

\section{Additional Evaluation on LLMSecEval}
\label{apx:llmseceval}
\paragraph{Setup.}
\final{In addition to the main evaluation on CWEval and BaxBench, we conduct a supplementary evaluation on LLMSecEval~\citep{tony2023llmseceval}, which consists of 150 natural-language programming tasks in Python and C. Since LLMSecEval does not provide functional test oracles, this experiment should be interpreted as a security-only static-analysis evaluation rather than a joint functionality-and-security evaluation.}
\paragraph{Metrics.}
\final{Security is assessed using CodeQL queries~\citep{CodeQLGitHub} and the ICD metric~\citep{cyberseceval,le2024indict}, both of which report CWE-specific vulnerabilities. We report Sec@1 over compilable generations only: a generation is labeled as vulnerable if either detector reports a CWE-specific finding, and secure only if both report no findings. Table~\ref{tab:icd_codeql_compare} summarizes the CWE-specific detection-rule coverage of each analysis tool. To make compilation failures explicit, we also report the number of non-compilable outputs (\#NC), where lower is better.}
\paragraph{Results.}
\final{Table~\ref{tab:llmseceval} shows that \sysname\ achieves the highest Sec@1 across all five evaluated models. The gains are especially pronounced for GPT-4o-mini and Gemini 2.5 Flash-Lite, where \sysname\ improves substantially over both direct prompting and prior security-oriented baselines. Importantly, these improvements do not come from producing fewer compilable programs: \sysname\ also maintains a low number of non-compilable outputs across models.}

\input{tables/apx/coverage_ls}

\input{tables/apx_ls}

\section{Prompts}\label{apx:prompts}
In this section, we provide prompts we used for the agents in \sysname\ design. We note that the prompt for the Planning agent is omitted here as we utilize the prompt from previous research~\cite{islam-etal-2025-codesim}.

% \begin{figure}
%     \centering
%     \includegraphics[width=\linewidth] {figures/apx/prompts/prompt_for_refinement_v2.pdf}
%    \caption{\revision{The prompt used for Knowledge Base Refinement to normalize raw text into a structured schema.}}
%     \label{fig:prompt_kb_refinement}
% \end{figure}

% \begin{figure*}
%     \centering
%     \includegraphics[scale=0.7,keepaspectratio]{figures/apx/prompts/prompt_figure_1.pdf}
%     \caption{The prompt used by the Security Advisor Agent to evaluate draft code and extract potential security risks.}
%     \label{fig:prompt_secadv1}
% \end{figure*}
\begin{figure*}
    \centering
    \includegraphics[width=0.97\textwidth]{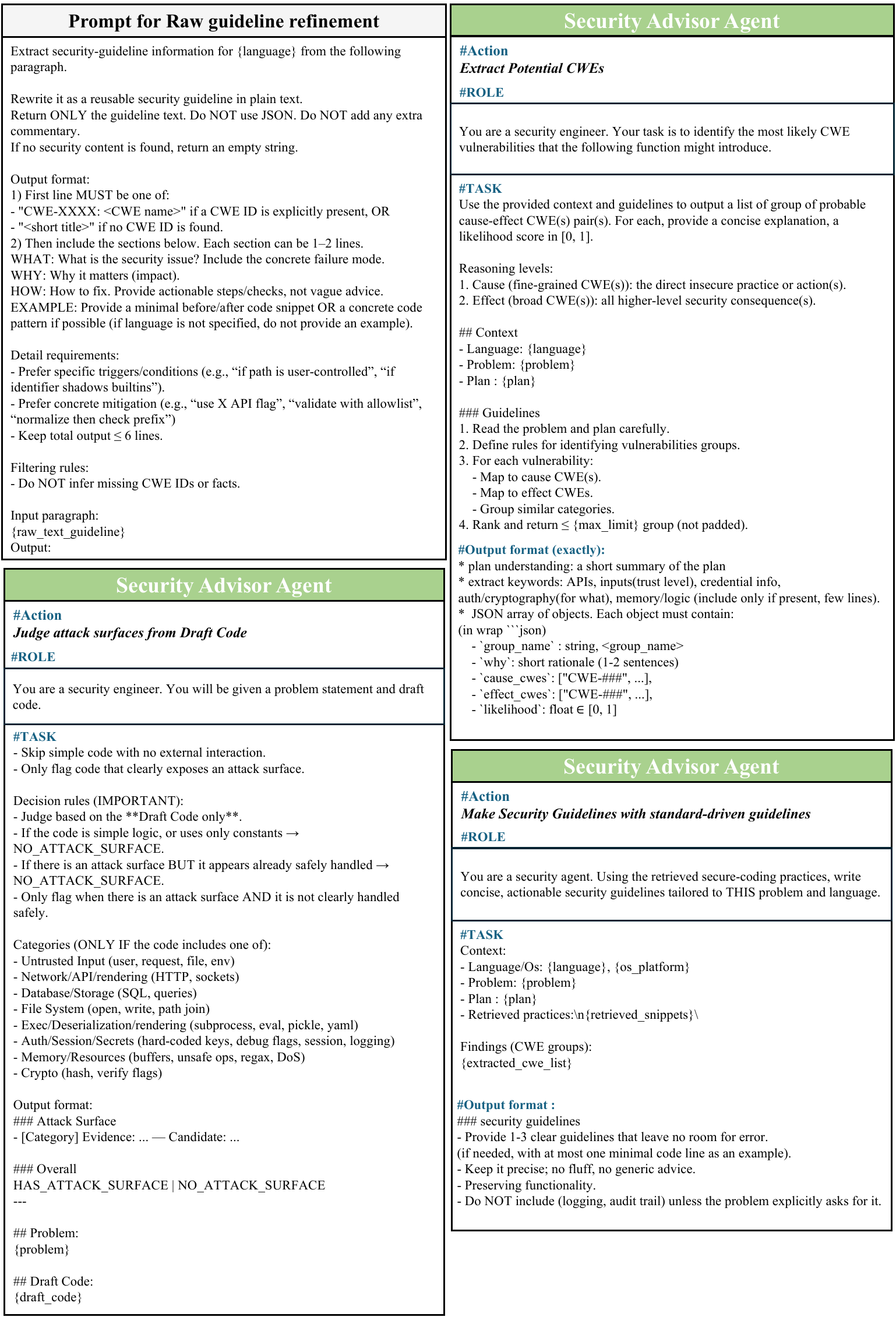}
    % \caption{\final{Prompts used by the Security Advisor Agent. The agent first performs an early-exit attack-surface check on the draft code, then extracts likely cause--effect CWE groups, and finally generates task-specific security guidelines from the retrieved secure-coding practices.}}
    \caption{ \final{Top-left: The prompt used for Knowledge Base Refinement to normalize raw text into a structured schema.} \final{Remaining three: Prompts used by the Security Advisor Agent. The agent first performs an early-exit attack-surface check on the draft code, then extracts likely cause--effect CWE groups, and finally generates task-specific security guidelines from the retrieved secure-coding practices.}}
    \label{fig:prompt_secadv1}
\end{figure*}

% \clearpage

\begin{figure*}    
\centering
    \includegraphics[width=\textwidth]{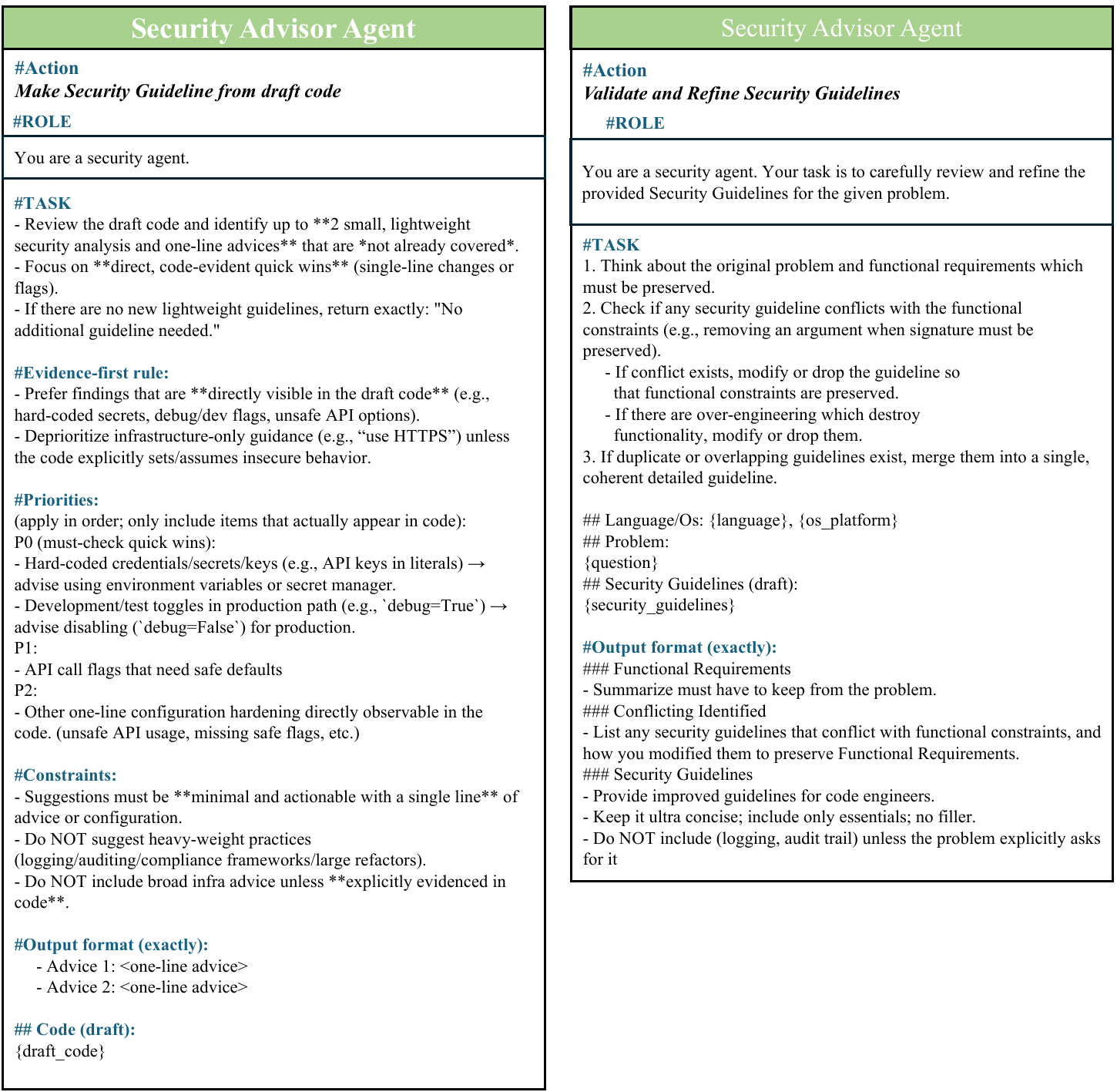}
    \caption{\final{Prompts used in the later stages of the Security Advisor pipeline. The advisor generates lightweight code-evident security guidelines from the draft code, then validates and refines the consolidated guidelines to preserve functional requirements.}}
    \label{fig:prompt_secadv2}
\end{figure*}
\clearpage

\begin{figure*}  
    \centering
    \includegraphics[width=\textwidth]{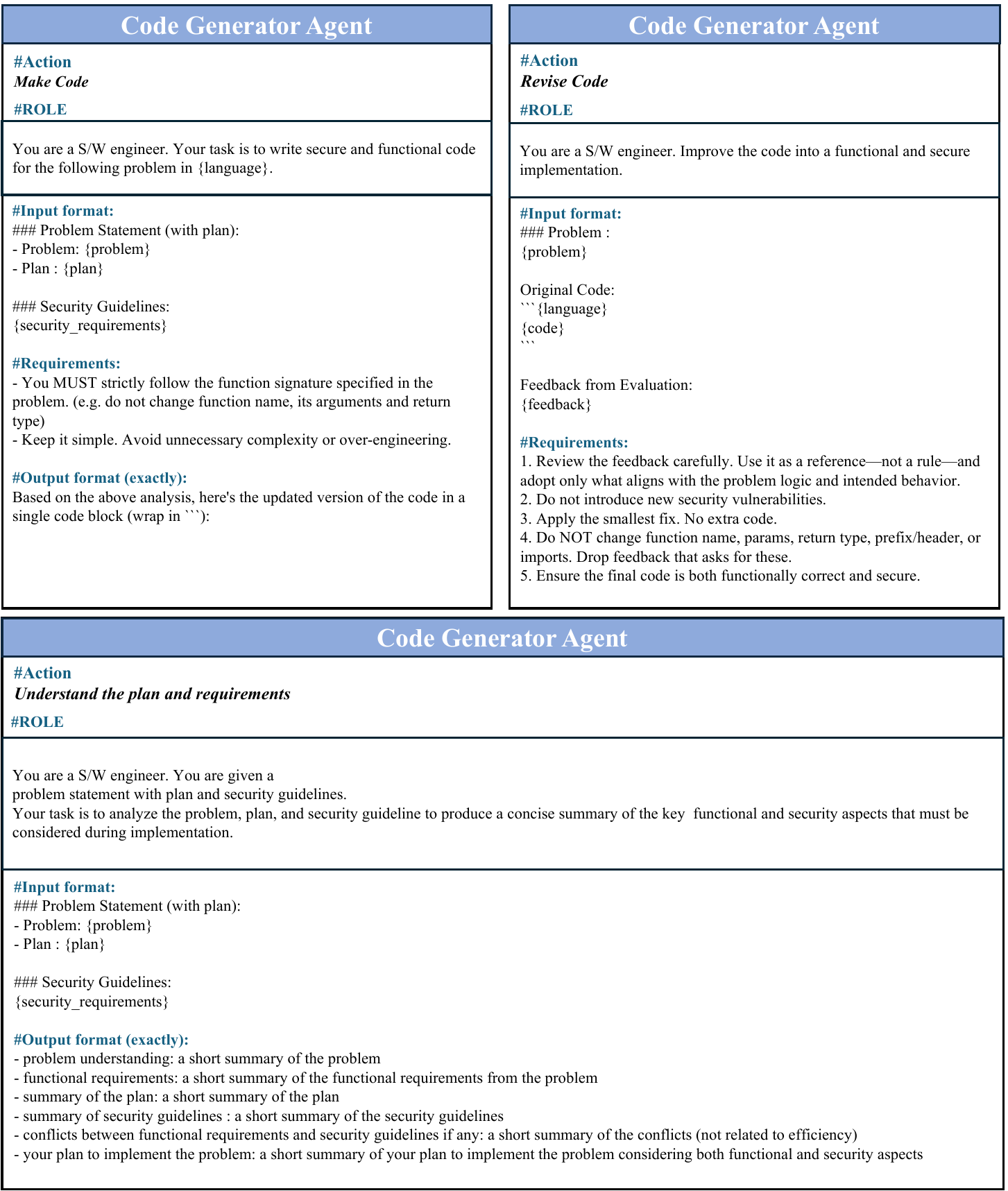}
    \caption{\final{Prompts used by the Code Generator Agent. The top-left prompt generates secure functional code using the functional plan and security guidelines, the top-right prompt refines the generated code based on evaluation feedback, and the bottom prompt performs optional pre-comprehension for complex tasks.}}
    \label{fig:prompt_secadv3_codegen}
\end{figure*}

\begin{figure*}[p] 
    \centering
    \includegraphics[width=0.98\textwidth,keepaspectratio]{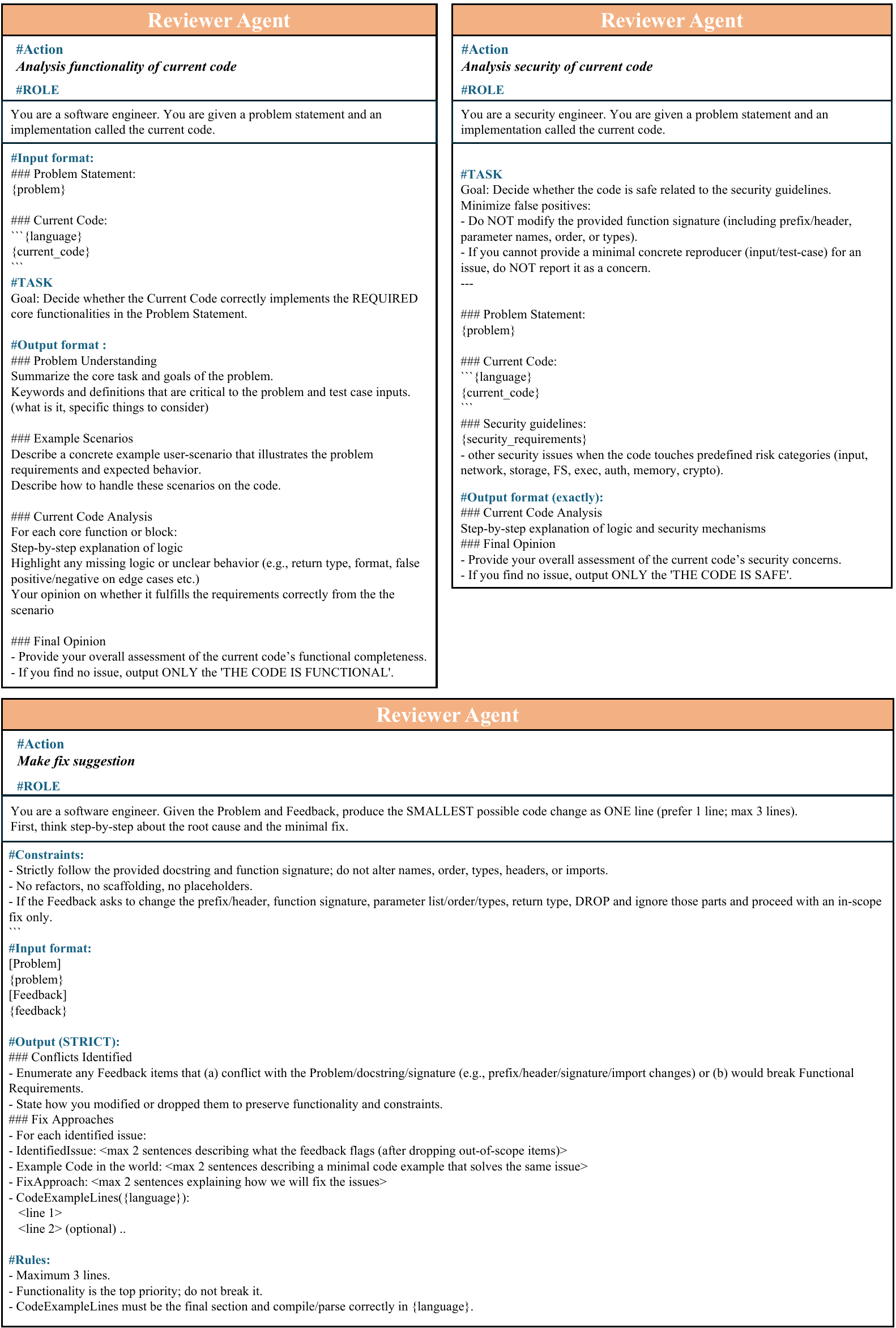}
    \caption{\final{Prompts used by the Reviewer Agent. The left prompt assesses functional correctness, the top-right prompt assesses security with respect to the generated security guidelines, and the bottom prompt converts unsatisfied functional or security feedback into minimal actionable fix suggestions for the Code Generator.}}
    \label{fig:prompt_secadv3_codegen}
\end{figure*}

%% file: tables/apx/effect_kb.tex
\begin{table}[htbp]
  \centering
  \small
  \resizebox{\linewidth}{!}{
  \setlength{\tabcolsep}{1.5mm}
  \begin{tabular}{clccc}
    \toprule
    \textbf{Model} &
    \textbf{Method} &
    \textbf{Func@1 (\%)} &
    \textbf{Sec@1 (\%)} &
    \textbf{F\&S@1 (\%)} \\
    \midrule
    \multirow{3}{*}{GPT-4o}
     & w/o guideline synthesis  & 76.47 & 67.83 & 57.98 \\
      & w/o knowledge-base refinement & 73.95 & 72.41 & 61.34 \\
      & \sysname\            & \final{79.83} & \final{79.31} & \final{\textbf{70.59}} \\
    \midrule
    \multirow{3}{*}{GPT-4o-mini}
    & w/o guideline synthesis        & 73.11 & 66.38 & 55.46 \\
      & w/o knowledge-base refinement & 65.66 & 76.72 & 56.30 \\
      & \sysname\  & \final{74.79} & \final{76.72} & \final{\textbf{66.39}} \\
    \bottomrule
  \end{tabular}}
  \caption{
  Impact of guideline synthesis and knowledge-base refinement on CWEval.
  }
  \label{tab:kb_refinement}
\end{table}

%% file: tables/apx/with_rag_k.tex
\begin{table}[!ht]
  \centering
  \small
  \setlength{\tabcolsep}{4mm}
  \renewcommand{\arraystretch}{0.8}
  \resizebox{\linewidth}{!}{
  \begin{tabular}{ccccc}
    \toprule
    \textbf{$C$} & \textbf{F@1} & \textbf{S@1} & \textbf{F\&S@1} & \textbf{Avg. API Cost(\$)} \\
    \midrule
    1 & \final{74.03} & \final{74.03} & \final{62.34} &  \final{0.034} \\ 
    \midrule
    3  & \final{83.12} & \final{73.33} & \final{67.53} &  \final{0.035} \\ 
    \midrule
     5 & \final{81.82} & \final{69.74} & \final{66.23} &  \final{0.036} \\ 
     \midrule
     10 & \final{84.42} & \final{74.67} & \final{68.83} &  \final{0.037} \\ 
    \bottomrule
  \end{tabular}}
\caption{\revision{CWEval results over 77 C/C++/Python tasks with GPT-4o under varying maximum CWE groups $C$.
Avg. API Cost (\$) denotes the average per-task GPT-4o API cost (USD).}}
  \label{tab:cwe_groups}
\end{table}

% [cwe1]
%   total entries     : 77
%   draft_pass=True   : 20
%   non-draft_pass    : 57
%   cwe_output counts : min=1, max=2, avg=1.02

% [cwe5]
%   total entries     : 77
%   draft_pass=True   : 19
%   non-draft_pass    : 58
%   cwe_output counts : min=2, max=4, avg=2.97
% 2개: 13건 (22%)
% 3개: 34건 (59%)
% 4개: 11건 (19%)
% 5개(최대): 0건
% [cwe10]
%   total entries     : 77
%   draft_pass=True   : 18
%   non-draft_pass    : 59
%   cwe_output counts : min=2, max=5, avg=3.63
% 3개: 25건 (42%)
% 4개: 25건 (42%)
% 5개: 7건 (12%)
% 6개 이상: 0건

%% file: tables/apx/early_exit_analysis.tex
\begin{table}[t]
  \centering
  \small
  \resizebox{\linewidth}{!}{
  \setlength{\tabcolsep}{1.2mm}
  {\renewcommand{\arraystretch}{0.85}
  \begin{tabular}{lcccc}
    \toprule
    \textbf{Model} & 
    \textbf{\# EE} &
    \textbf{\# Not Secure} &
    \textbf{Not Secure / EE (\%)} &
    \textbf{Not Secure / Total (\%)} \\
    \midrule
    GPT-4o                   & 30 & 2 &  6.7 & 1.7 \\
    GPT-4o-mini              & 13 & 3 & 23.1 & 2.5 \\
    Gemini 2.5 Flash         & 29 & 2 &  6.9 & 1.7 \\
    Gemini 2.5 Flash-lite    & 17 & 2 & 11.8 & 1.7 \\
    DeepSeek-R1-Distill-70B  & 55 & 4 &  7.3 & 3.4 \\
    Qwen3-8B                 & 45 & 6 & 13.3 & 5.0 \\
    \bottomrule
  \end{tabular}}}
  \caption{\final{Early-exit triage analysis on CWEval.
  \# EE denotes the number of tasks where the Security Advisor
  exits without full review. \# Not Secure counts EE cases
  subsequently judged as functionally correct but insecure.
  Not Secure / EE measures this rate among exited tasks;
  Not Secure / Total measures it across the full benchmark.}}
  \label{tab:early_exit_audit}
\end{table}

%% file: tables/apx/shared.tex
\begin{table}[t]
  \centering
  \small
  \resizebox{\linewidth}{!}{
  \setlength{\tabcolsep}{1.2mm}
  {\renewcommand{\arraystretch}{0.8}
  \begin{tabular}{clccc}
    \toprule
    \textbf{Model} & 
    \textbf{Method} &
     \textbf{Func@1 (\%)} &
     \textbf{Sec@1 (\%)} &
     \textbf{F\&S@1 (\%)} \\
    \midrule

    % ===== GPT-4o =====
    \multirow{2}{*}{GPT-4o}
    & \sysname-Shared  & 79.83 & 77.59 & 68.91 \\
      & \sysname\          & 79.83 & 79.31 & 70.59 \\
    \midrule

    % (placeholder for another model if needed)
    \multirow{2}{*}{GPT-4o-mini}
    & \sysname-Shared & 52.94  & 66.67 & 46.22 \\
    & \sysname\ & 74.79 & 76.72 & 66.39   \\
    \bottomrule
  \end{tabular}}}
  
\caption{
\final{Comparison of \sysname\ variants on CWEval.}
}

  \label{tab:shared_cweval}
\end{table}

%% file: tables/apx/shared_bax.tex
\begin{table}[t]
  \centering
  \small
  \resizebox{\linewidth}{!}{
  \setlength{\tabcolsep}{3.0mm}
  {\renewcommand{\arraystretch}{0.8}
  \begin{tabular}{clccc}
    \toprule
    \textbf{Model} & 
    \textbf{Method} &
     \textbf{Func@1 (\%)} &
     \textbf{F\&S@1 (\%)} \\
    \midrule

    % ===== GPT-4o =====
    \multirow{2}{*}{GPT-4o}
      & \sysname-Shared  & 39.80  & 22.45 \\
      & \sysname\          & 49.74  & 31.12 \\
    \midrule

    % (placeholder for another model if needed)
    \multirow{2}{*}{GPT-4o-mini}
    & \sysname-Shared & 27.04  & 19.64 \\
    & \sysname\ & 29.59  & 21.94   \\
    \bottomrule
  \end{tabular}}}
  
\caption{
\final{Comparison of \sysname\ variants on BaxBench.}
}

  \label{tab:shared_bax}
\end{table}

%% file: tables/apx/coverage_ls.tex
\begin{table}[!ht]
  \centering
  \small
  \setlength{\tabcolsep}{4mm}
  \renewcommand{\arraystretch}{0.5}
  \resizebox{0.9\linewidth}{!}{
  \begin{tabular}{llccc}
    \toprule
    \textbf{Analyzer} & \textbf{Metric} & \textbf{C/C++} & \textbf{Python} \\
    \midrule
    \multirow{2}{*}{ICD} 
      & \# of rules & 34 & 24  \\ 
      & \# of CWEs & 19 & 9  \\ 
    \midrule
    \multirow{2}{*}{CodeQL} 
      & \# of *.ql & 54 & 45  \\ 
      & \# of CWEs & 34 & 31  \\ 
    \bottomrule
  \end{tabular}}
\caption{Comparison of scanning rule counts and CWE coverage across programming languages under ICD and CodeQL configurations.}
  \label{tab:icd_codeql_compare}
\end{table}

%% file: tables/apx_ls.tex
\begin{table}[!ht]
  \centering
  \footnotesize
  \setlength{\tabcolsep}{3.5mm} % 열 간격 적당히
  \renewcommand{\arraystretch}{0.90} % 행 간격 압축
  \resizebox{\linewidth}{!}{
  % \begin{tabular}{@{}clccccc@{}}
  \begin{tabular}{clcc}
    \toprule
    \textbf{Model} & \textbf{Method} 
    & \textbf{Sec@1 (\%) $\uparrow$} 
    & \textbf{\#NC $\downarrow$} \\
    \midrule

    % ===== GPT-4o =====
    \multirow{6}{*}{GPT-4o}
      & Direct        & 44.00 & 0 \\
      & \textsc{SecGuide}   & 78.47 & 6 \\
      & \revision{\textsc{CodeGuarder}}  & \revision{62.25} & 9 \\
      & \textsc{INDICT}    & 51.68 & 1 \\
      & \textsc{RESCUE}    & 50.68 & 2 \\
      & \sysname\   & \textbf{79.19} & 1 \\
    \midrule

    % ===== GPT-4o-mini =====
    \multirow{6}{*}{GPT-4o-mini}
      & Direct       & 43.15 & 4 \\
      & \textsc{SecGuide}  & 70.63  & 7 \\
      & \revision{\textsc{CodeGuarder}}  & \revision{58.33} & \revision{6} \\
      & \textsc{INDICT}   & 54.36 & 1 \\
      & \textsc{RESCUE}   & 42.07 & 5 \\
      & \sysname\  & \textbf{80.54} & 1 \\
    \midrule

    % ===== Gemini 2.5 Flash =====
    \multirow{6}{*}{\makecell{Gemini 2.5\\Flash}}
      & Direct        & 44.59 & 2 \\
      & \textsc{SecGuide}  & 51.06 & 9 \\
      & \revision{\textsc{CodeGuarder}}  & \revision{66.21} & \revision{5} \\
      & \textsc{INDICT}    & 52.14 & 10 \\
      & \textsc{RESCUE}   & 56.34 & 8 \\
      & \sysname\   & \textbf{70.00} & \revision{0} \\
    \midrule

    % ===== Gemini 2.5 Flash-Lite =====
    \multirow{6}{*}{\makecell{Gemini 2.5\\Flash-Lite}}
      & Direct       & 49.66  & 1 \\
      & \textsc{SecGuide}  & 51.11 & 15 \\
      & \revision{\textsc{CodeGuarder}}  & \revision{53.57} & \revision{10} \\
      & \textsc{INDICT}    & 60.96 & 4 \\
      & \textsc{RESCUE}   & 44.76  & 7 \\
      & \sysname\   & \textbf{78.52} &\revision{1} \\
    \midrule

    % ===== DeepSeek-R1 (70B) =====
    \multirow{6}{*}{\revision{\makecell{DeepSeek\mbox{-}R1\\ \mbox{-}Distill (70B)}}}
      & Direct       & 53.85  & 7 \\
      & \textsc{SecGuide} & 54.20 & 19 \\
      & \revision{\textsc{CodeGuarder}} & \revision{71.22} & \revision{11} \\
      & \textsc{INDICT}   & 53.57  & 10 \\
      & \textsc{RESCUE}   & 53.79  & 18 \\
      & \sysname\   & \revision{\textbf{72.11}} & 3 \\
    \bottomrule
  \end{tabular}
  }
% DeepSeek-R1-Distill (70B)
\caption{
\final{Evaluation on LLMSecEval (150 tasks, two languages). \#NC counts non-compilable cases (lower is better).}
}

  \label{tab:llmseceval}
\end{table}

% $\Delta$ indicates relative improvement over \emph{Direct}, and 